#\documentclass[manuscript]{aastex6}

\documentclass[onecolumn]{aastex6}

\newcommand{\Lya}{Ly$\alpha$\ }

\def\OIIg{\ifmmode {{\rm [OII]\lambda\lambda 3726,3729}}\else
                ${\rm [OII]\lambda\lambda3726,3729}$\fi}

\def\SIIg{\ifmmode {{\rm [SII]\lambda\lambda6717,6731}}\else
                ${\rm [SII]\lambda\lambda6717,6731}$\fi}
                
\def\SII{\ifmmode {{\rm [SII]\lambda}}\else
                ${\rm [SII]\lambda}$\fi}

\def\Hag{\ifmmode {{\rm H\alpha\ \lambda6563+[NII]\lambda\lambda6548,6584}}\else
                ${\rm H\alpha\ \lambda6563+[NII]\lambda\lambda6548,6584}$\fi}

\def\NII{\ifmmode {{\rm [NII]\lambda}}\else
                ${\rm [NII]\lambda}$\fi}

\def\Ha{\ifmmode {{\rm H\alpha\ \lambda}}\else
                 ${\rm H\alpha\ \lambda}$\fi}

\def\OIIIg{\ifmmode {{\rm [OIII]\lambda\lambda4959\AA,5007\AA}}\else
                ${\rm [OIII]\lambda\lambda4959,5007}$\fi}

\def\Hb{\ifmmode {{\rm H\beta\ \lambda}}\else
                ${\rm H\beta\ \lambda}$\fi}

\def\OIII{\ifmmode {{\rm [OIII]\lambda}}\else
                ${\rm [OIII]\lambda}$\fi}                
           
\begin{document}
\title{A Two-Dimensional Spectroscopic Study of Emission-Line Galaxies in the Faint Infrared Grism Survey (FIGS). I: Detection Method and Catalog}

\author{Norbert Pirzkal\altaffilmark{1}}
\author{Barry Rothberg\altaffilmark{2,3}}
\author{Russell E. Ryan\altaffilmark{1}}
\author{Sangeeta Malhotra\altaffilmark{4,5}}
\author{James Rhoads\altaffilmark{4,5}}
\author{Norman Grogin\altaffilmark{1}}
\author{Emma Curtis-Lake\altaffilmark{6}}
\author{Jacopo Chevallard\altaffilmark{6}}
\author{Stephane Charlot\altaffilmark{6}}
\author{Steven L. Finkelstein\altaffilmark{7}}
\author{Anton M. Koekemoer\altaffilmark{1}}
\author{Parviz Ghavamian\altaffilmark{8}}
\author{Myriam Rodrigues\altaffilmark{9}}
\author{Fran\c{c}ois Hammer\altaffilmark{9}}
\author{Mathieu Puech\altaffilmark{9}}
\author{Rebecca L. Larson\altaffilmark{7}}
\author{Lise Christensen\altaffilmark{10}}
\author{Andrea Cimatti\altaffilmark{11,12}}
\author{Ignacio Ferreras\altaffilmark{13}}
\author{Jonathan P. Gardner\altaffilmark{5}}
\author{Caryl Gronwall\altaffilmark{14,15}}
\author{Nimish P. Hathi\altaffilmark{1}}
\author{Bhavin Joshi\altaffilmark{4}}
\author{Harald Kuntschner\altaffilmark{16}}
\author{Gerhardt R. Meurer\altaffilmark{17}}
\author{Robert W. O'Connell\altaffilmark{18}}
\author{Goeran Oestlin\altaffilmark{19}}
\author{Anna Pasquali\altaffilmark{20}}
\author{John Pharo\altaffilmark{2}}
\author{Amber N. Straughn\altaffilmark{5}}
\author{Jeremy R. Walsh\altaffilmark{16}}
\author{Darach Watson\altaffilmark{10}}
\author{Rogier A. Windhorst\altaffilmark{2}}
\author{Nadia L Zakamska\altaffilmark{21}}
%\author{Andrew Zirm\altaffilmark{22}}

\altaffiltext{1}{Space Telescope Science Institute, Baltimore, MD 21210, USA}
\altaffiltext{2}{Large Binocular Telescope Observatory,University of Arizona, AZ 85721, USA}
\altaffiltext{3}{George Mason University, Fairfax, VA 22030, USA}
\altaffiltext{4}{Arizona State University, School of Earth and Space Exploration, Tempe, AZ 85287, USA}
\altaffiltext{5}{NASA’s Goddard Space Flight Center, Astrophysics Science Division, Code 660, Greenbelt MD 20771, USA}
\altaffiltext{6}{Sorbonne Universite\'s, UPMC-CNRS, UMR7095, Institut d’Astrophysique de Paris, F-75014, Paris, France}
\altaffiltext{7}{University of Texas at Austin, Austin, TX 78712, USA}

\altaffiltext{8}{Department of Physics, Astronomy and Geosciences, Towson University, Towson, MD 21252, USA}

\altaffiltext{9}{GEPI, Observatoire de Paris, PSL University, CNRS, 5 Place Jules Janssen, F-92190 Meudon, France}

\altaffiltext{10}{Dark Cosmology Centre, Niels Bohr Institute, University of Copenhagen, DK-2100, Denmark} %, Juliane Maries Vej 30, DK-2100 Copenhagen, Denmark}
\altaffiltext{11}{University of Bologna, Department of Physics and Astronomy, Via Gobetti 93/2, I-40129, Bologna, Italy} 
\altaffiltext{12}{INAF - Osservatorio Astrofisico di Arcetri,
Largo E. Fermi 5, I-50125, Firenze, Italy}
\altaffiltext{13}{Mullard Space Science Laboratory, University College London, Holmbury St Mary, Dorking, Surrey, RH5 6NT, UK}
\altaffiltext{14}{Department of Astronomy and Astrophysics, The Pennsylvania State University, University Park, PA 16802, USA }
\altaffiltext{15}{ Institute for Gravitation and the Cosmos, The Pennsylvania State University, University Park, PA 16802, USA}
\altaffiltext{16}{European Southern Observatory, D-85748, Garching, Germany}
\altaffiltext{17}{International Centre for Radio Astronomy Research, The University of Western Australia, Crawley, WA 6009, Australia}
\altaffiltext{18}{The University of Virginia, Charlottesville, VA 22904-4325, USA }
\altaffiltext{19}{Stockholm University, SE-10691, Stockholm, Sweden}
\altaffiltext{20}{Astronomisches Rechen-Institut, Zentrum f{\"ur} Astronomie der Universit{\"a}t Heidelberg, D-69120, Heidelberg, Germany}
\altaffiltext{21}{Department of Physics and Astronomy, Johns Hopkins University, Baltimore, MD 21218, USA
} %, 3400 N. Charles St., Baltimore MD 21218}
%\altaffiltext{22}{University of Copenhagen, Niels Bohr Institute, København, DK-2100, Denmark}

%\maketitle
\begin{abstract}
% \Lya \\
% \OIIg\\
% \SIIg\\
% \Hag\\
% \OIIIg\\
% \Hb\\
% \Lya\\
% \Hb4861\\
% \OIII4959\\
% \OIII5007\\

We present the results from the application of a two-dimensional emission line detection method, EMission-line two-Dimensional (EM2D), to the near-infrared G102 grism observations obtained with the Wide-Field Camera 3 (WFC3) as part of the Cycle 22 {\em Hubble Space Telescope} Treasury Program:  the Faint Infrared Grism Survey (FIGS). Using the EM2D method, we have assembled a catalog of emission line galaxies (ELGs) with resolved star formation from each of the four FIGS fields.  Not only can one better assess the  global properties of ELGs, but the EM2D method allows for the analysis and an improved study of  
the individual emission-line region {\it within} each galaxy.  This paper includes a description of the methodology, advantages, and the first results of the EM2D method applied to ELGs in FIGS.  The advantage of 2D emission line measurements includes significant improvement of galaxy redshift measurements, approaching the level of accuracy seen in high-spectral-resolution data, but with greater efficiency; and the ability to identify and measure the properties of multiple sites of star-formation and over scales of $\sim$ 1 kpc within individual galaxies out to z $\sim$ 4. The EM2D method also significantly improves the reliability of high-redshift ($z\sim7$) Lyman-$\alpha$ detections.  Coupled with the wide field of view and high efficiency of space-based grism observations, EM2D provides a noteworthy improvement on the physical parameters that can be extracted from grism observations.

\end{abstract}
\section{Introduction}
One indicator of (relatively) recent star-formation in galaxies is the presence of strong emission lines  \citep[see][and references therein]{Kennicutt1998}. Recombination lines, such as Ly$\alpha$, H$\alpha$, H$\beta$\ in the rest-frame UV and optical, and the prominent rest-frame optical forbidden emission lines [OII] and [OIII] are all integral in tracing the ionizing flux produced by short lived ($\sim$10 Myr), massive (${ > 10 \ M_\sun}$) stars. It is important to note that these lines are produced not only in the central nuclear regions, but in star-forming regions throughout galaxies.   Emission-line galaxies (ELGs) are predominantly identified in narrowband photometric \citep[e.g.][]{Djorgovski1985, Boroson1993} or spectroscopic grism surveys \citep[e.g.][]{Mayall1936, Markarian1967, Smith1975}.  ELGs are easily identifiable because a significant amount of the photons radiated by them originate in star-forming regions producing strong emission lines. More recently, with the availability of slitless grism modes on the Hubble Space Telescope (HST), with the Near Infrared Camera and Multi-Object Spectrometer (NICMOS), the Advanced Camera for Surveys (ACS), the Wide-Field Camera 3 (WFC3), several projects, such as NICMOS/HST Grism Parallel Survey \citep{McCarthy1999}, the ACS Pure Parallel Ly$\alpha$ Emission Survey \citep[APPLES;][]{Pasquali2003}, the Grism ACS Program for Extragalactic Science \citep[GRAPES;][]{Pirzkal2004}, Probing Evolution And Reionization Spectroscopically \citep[PEARS;][]{Pirzkal2009}, the WFC3 Infrared Spectroscopic Parallel Survey \citep[WISP;][]{Atek2010}, the Grism Lens-Amplified Survey from Space \citep[GLASS;][]{Treu2015}, 3D-HST \citep{Momcheva2016}, and the Faint Infrared Galaxy Survey \citep[FIGS;][]{Pirzkal2017} have identified a large population of star-forming galaxies (SFGs) over a nearly contiguous range of redshifts ($0<z<3.5)$, including the epoch of peak star formation at $1.5 < z < 2.5$ \citep{Madau1998,Hopkins2004,Madau2014}. 

The presence of bright, easily identifiable emission lines makes the spectroscopic determination of the redshift of these individual ELGs straightforward. These same emission lines also allow for the direct measurement of physical properties, such as star formation rates, ages, and metallicities, of both the star forming regions within ELGs, as well as inferring the overall global properties of galaxies \citep[e.g.][]{Aller1942,McGaugh1991,Kewley2001,Nagao2006}. This includes variations in kinematics and star-forming properties across a galaxy \citep[e.g.][]{Rubin1970,Rubin1972}.

For the most part, space-based HST grism surveys have focused primarily on the {\it integrated} properties of SFGs and ELGs. These programs have preferred to sample wider angular coverage on the sky or random parallel fields in order to increase the potential sample size and mitigate cosmic variance. The downside is that such surveys often use only one or two orientations on the sky in an effort to increase efficiency and reduce overheads. Yet, this method also increases the effects and the impact of contamination from other sources in or near the field of view (i.e. light dispersed from foreground objects in the field, or just outside the field, which contaminates the dispersed light of targets of interest).  Most surveys also sacrifice depth per field in favor of larger angular sky coverage.  While space-based slitless grism observations are often more effective than ground-based counterparts, particularly at longer optical and near-infrared wavelengths (where telluric emission and absorption lines reduce any sensitivity gains from larger aperture mirrors), they still must be carefully planned to avoid contamination from non-telluric sources.

In an earlier series of papers \citep{Straughn2008, Straughn2009, Pirzkal2013}, we successfully demonstrated how ACS slitless grism observations obtained using multiple orientations on the sky, could be used to significantly improve the wavelength accuracy of emission lines, as well as actually identifying multiple emission line sources {\it within} a galaxy by avoiding self-contamination (i.e. the dispersed light of one part of the galaxy, contaminating the dispersed light from an adjoining part of the galaxy when using a single orientation on sky).  However, our earlier series of papers were restricted to the redshift range of $0 < z < 1.5$\ using rest-frame H$\alpha$, [OIII], or [OII] obtained with ACS on HST. While the results were interesting, they fell just short of probing the important peak of star-formation ($1.5 < z < 2.5$).  

The use of the WFC3 camera and its near-IR grisms offers a comparably large field of view and flux sensitivities as the ACS grism, while allowing us to detect ELGs at redshifts up to $z=2$. The FIGS program obtained deep G102 observations of four distinct fields (two in GOODS-N and two in GOODS-S), each using five distinct orientations on sky \citep{Pirzkal2017}. While the choice of multiple position angles was driven primarily by the need to mitigate the amount of contamination by other dispersed nearby sources, it also created an advantageous situation that allowed to us to search for ELGs using the methods laid out in \cite{Pirzkal2013}, namely detecting the presence of emission lines and their source star-formation regions without having to first find or detect the underlying host galaxy.  The results from our work at $0 < z < 1.5$\ in PEARS \citep{Pirzkal2013} resulted in three key findings:  (1) the computed line luminosities showed evidence of a flattening in the luminosity function slope with increasing redshift; (2) the star-forming systems showed evidence of complex morphologies with star formation occurring predominantly within one effective (half-light) radius. However, the morphologies showed no correlation with host stellar mass; and (3) the number density of SFGs with $M_* \gg 10^{9} M_{\odot}$ decreases by an order of magnitude at $z \le 0.5$ relative to the number at $0.5 < z < 0.9$, supporting the argument of galaxy downsizing \citep{Cowie1996}.

% \section{Paper Outline}
% \begin{itemize}
%     \item Introduction
%     \item Description of EM2D
%     \item Applying EM2D to the FIGS sample
%     \item Description of emission line regions and emission line fluxes catalog
%     \item Accuracy of the redshifts obtained using the EM2D method and FIGS data
%     \item How complete is our emission line catalog, as a function of line luminosity and redshift (for Ha OIII and OII separately)
%     \item luminosity function of Ha OIII and OII and comparison the PEARS and literature (evolution?)
% \end{itemize}

In this paper, we probe up to and including the peak epoch of star-formation using near-IR observations obtained with the G102 grisms using the WFC3/IR camera.  First, we discuss the EM2D methodology, then apply it to the WFC3/IR G102 science data in  Section \ref{sec:EM2D_intro}.  Section \ref{EM2D:ID} explores emission line identification, and Section \ref{sec:FIGS_MCMC} describes how line fluxes are measured. Section \ref{EM2D:loc} describes the physical location of star forming regions in individual galaxies. Section \ref{sec:EM2D_ext} shows how we identified a significant number of galaxies with diffuse emission and how emission line maps can be created. In Section \ref{sec:completeness} we provide a discussion of the completeness of our EM2D survey. Finally, Sections \ref{sec:results_naked}, \ref{sec:results_highz} and \ref{sec:lum} discuss high-EW galaxies, high-redshift galaxies, and the luminosity functions we derive for the FIGS EM2D galaxy sample.
The goal of this paper is to demonstrate the effectiveness of two-dimensional emission line detection, diagnostics, and their application to discerning new insights into the key epoch of star-formation in the universe.
All calculations in this paper assume $H_0 = 67.3\ km\ s^{-1} Mpc^{-1}$\ and $\Omega_M = 0.315$, $\Omega_\Lambda = 0.685$ \citep{Planck2015}. All magnitudes are given in the $AB$\ system \citep{Oke1983}.

\section{Observations}\label{sec:Observations}
The data presented here are from the HST Cycle 22 Treasure Program Faint Infrared Galaxies (FIGS, Proposal ID: 13779, PI: S. Malhotra), which was awarded 160 orbits ($\sim 100$\  ks total exposure time) with the WFC3/IR instrument using the G102 grism filter.  The field of view of the WFC3/IR channel is 2''2 $\times$ 2''2 and the G102 grism has a resolution of 25\AA\ per pixel ($\approx36\AA$\ effective resolution since the WFC3 PSF is $\approx1.5$\ pixel wide). FIGS focuses on four  distinct fields in the Great Observatories Origins Deep Survey (GOODS) North and South fields \citep{Giavalisco2004}: two2 fields in GOODS-North and 2 fields in GOODS-South \citep[for details, see][]{Pirzkal2017}.  Each field was observed with five distinct position angles (PAs) in order to minimize contamination and maximize the signal-to-noise ratio (S/N).  The FIGS survey reaches a 3$\sigma$\ continuum depth of $\approx26$\ $AB$\ magnitudes and probes emission lines down to ${\rm \approx 10^{-17}\ erg\ s^{-1}\ cm^{-2}}$\ \citep{Pirzkal2017}.

\section{Data Reduction snd Analysis}
\subsection{The Emission-line 2D (EM2D) Method}\label{sec:EM2D_intro} 
\subsubsection{Methodology}\label{sec:EM2D_basic} 
\begin{figure}[!h]
    \centering
    \includegraphics[width=6in]{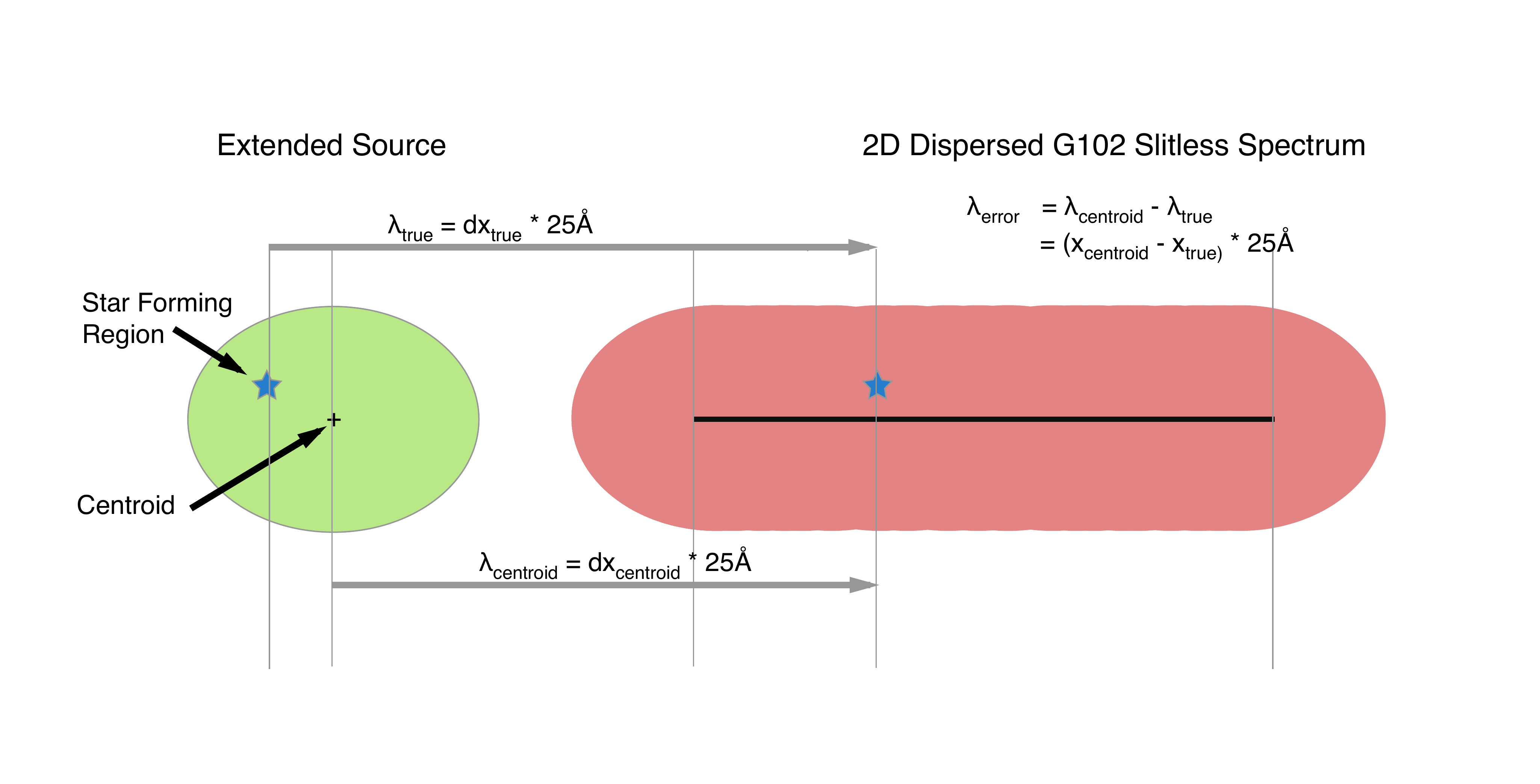}
    \caption{Illustration of how the difference in position between the true and estimated location of the origin of an emission line leads to an error in the observed wavelength estimate for that emission line.}
    \label{fig:EM2Da}
    
\end{figure}

In ``classical'' slit (or multi-slit) spectroscopy, only the the light from the object within the slit or slits is dispersed across the detector. While this avoids contamination from other nearby objects in the field, it reduces the survey efficiency significantly.  Slitless grism spectroscopy is more effective in terms of the number of objects for which light can be dispersed, and it is ideally suited for survey work in which one does not know a priori which objects in the field are of particular interest.  However, in a grism survey, each astronomical source acts as a dispersing object, potentially contaminating itself and other sources.  Moreover, light from higher grism orders can also be dispersed onto the detector, allowing each object to contaminate many other objects many times, even if the source is outside the nominal field of view as projected onto the detector.  Furthermore, a resolved object can ``self-contaminate,'' that is, it's own dispersed light falls on top of adjacent pixels in the target, which are themselves dispersing light.  At best, this results in the morphology of the source in the dispersion direction acting similar to a convolution kernel, which blurs the final spectrum in the dispersion direction. At worst, if the object is spectrally inhomogeneous, as in cases of an SFG with knots and clumps containing emission lines, determining the exact wavelength of the emission line requires detailed knowledge of where the signal originates within the object. Traditionally, both slit and slitless spectroscopy assume that light is produced at the center of the source and that the actual shape of the source has the same effect as a simple convolution kernel. While valid for slit spectroscopy (assuming the source is not spatially resolved {\it within} the slit), this assumption is not valid for slitless spectroscopy in which objects are resolved (even partially).  Figure \ref{fig:EM2Da} demonstrates that this assumption can result in a systematic error in the determination of the observed wavelength of an emission (or absorption) line from a region that is not centrally located (i.e. a star-forming knot, star-cluster, second nucleus in an unrelaxed merger, etc.) within a galaxy. 

\begin{figure}[!h]
    \centering
    \includegraphics[width=6in]{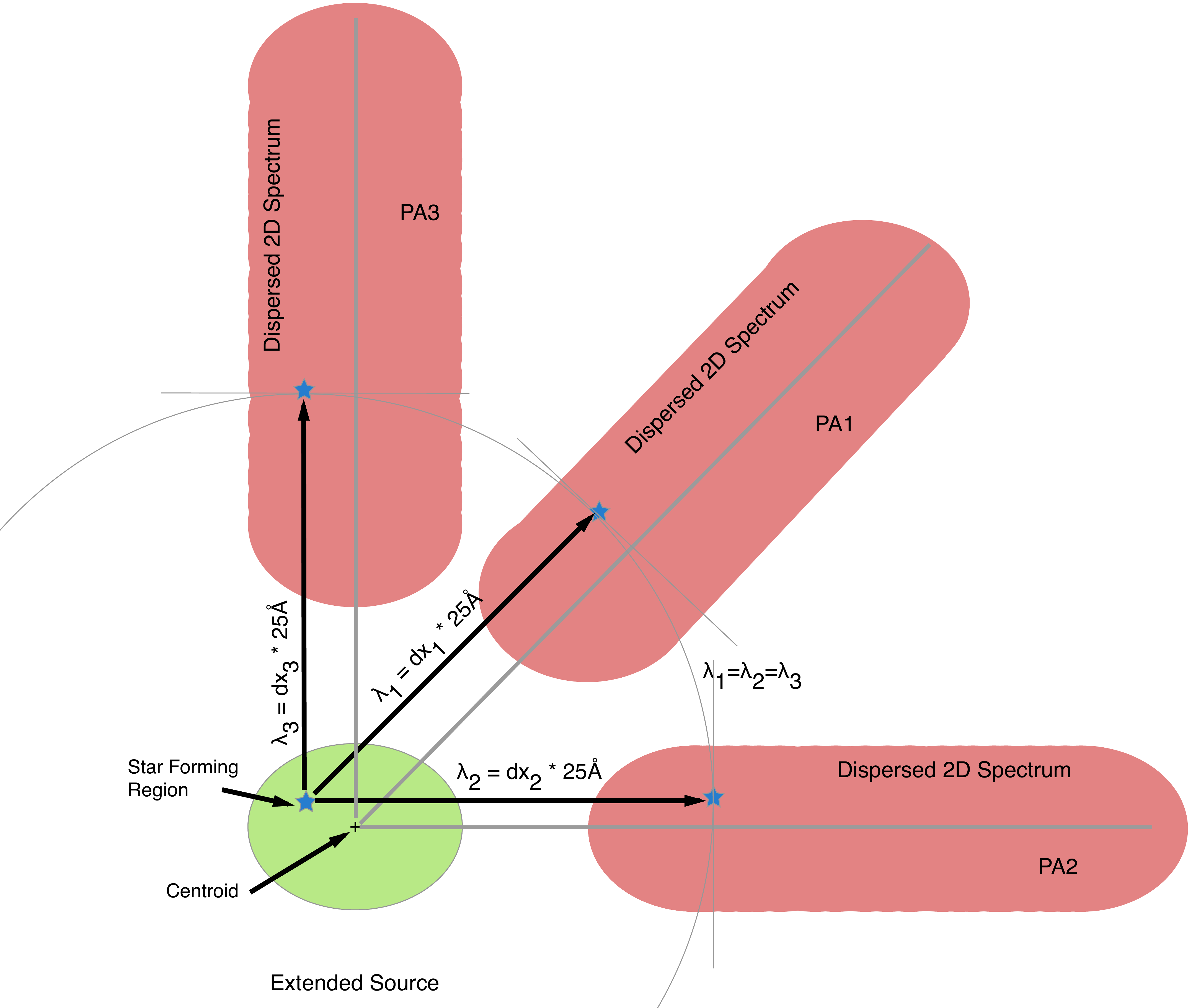}
    \caption{Figure illustrating the EM2D method and how multiple orientations allow us to both pinpoint the origin of an emission line as well as determine an accurate estimate of the observed wavelength of the emission line.}
    \label{fig:EM2Db}
\end{figure}

\begin{figure}[!h]
     \centering
     \includegraphics[width=7in]{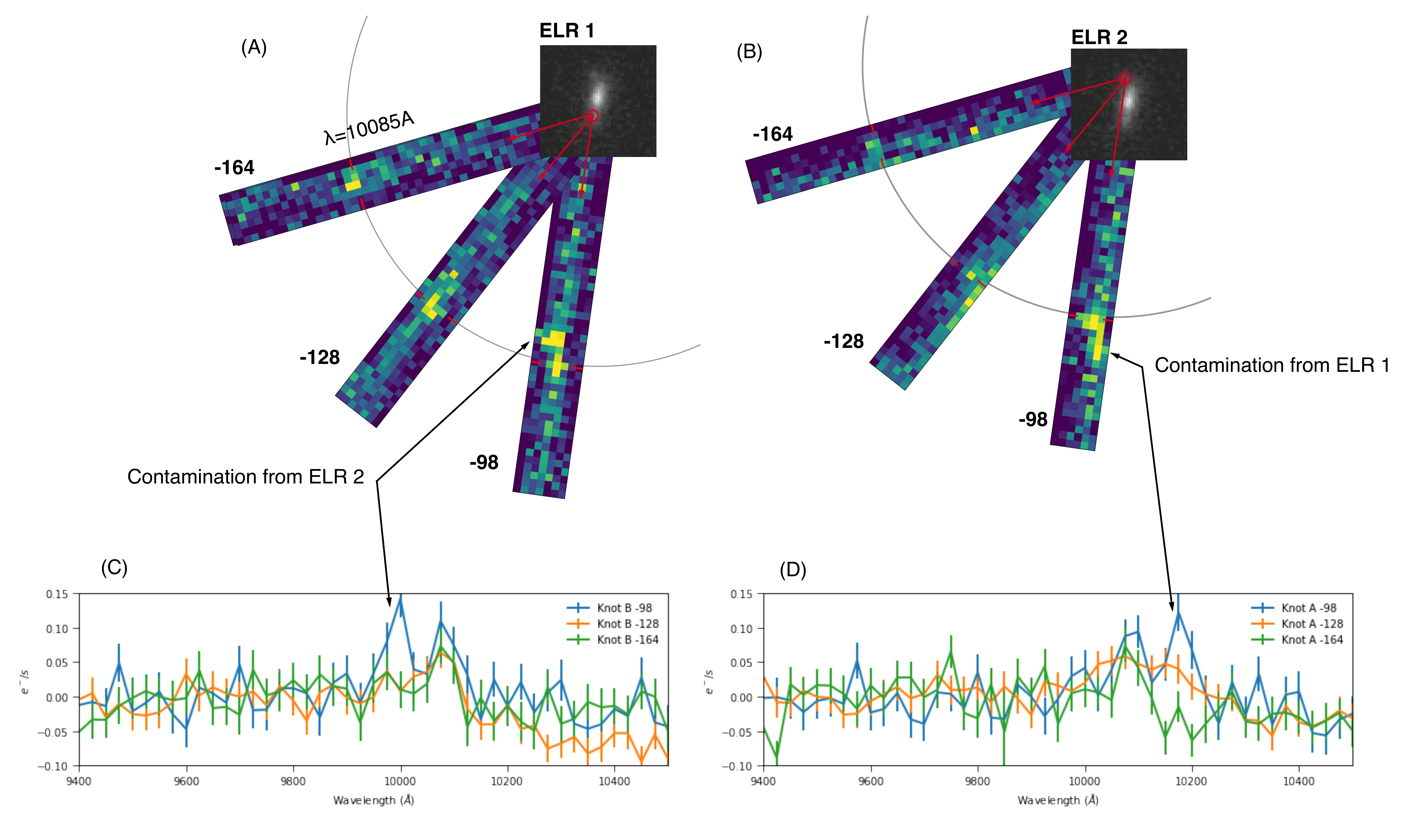}
     \caption{The EM2D method used in this paper. We rely on the detection of emission line candidates in dispersed images in different PAs and use our knowledge of the dispersion to determine the location of the emission-line regions. As this Figure illustrates, objects with multiple emission line regions (ELR1 and ELR2) can be identified, which is something not readily done using normal extraction. The latter can in fact lead to  erroneous line detection when multiple emission lines are produced within different emission line regions. Panels A and B show the location of the emission line for each ELR in the original observations, which is at a wavelength of $10085\AA$. The emission lines are marked on the 2D spectra using red lines and the dispersion directions are shown by red arrows pointing away from the source. Panels C and D show the extracted spectra and illustrate how different ELRs in a galaxy can contaminate each other. This is readily visible in the 2D spectra as well as the 1D spectra for PA=-98. On the other hand, PA=-164 allows for a clean extraction of the spectra, while PA=-128 results in some smearing of the emission line.}
     \label{fig:EM2D}
 \end{figure}

In a case where a component is resolved within a galaxy, rotating the field (i.e. changing the HST PAs) and dispersing the spectrum of the source in a different direction  projected on the sky allows one to recover {\it both} the exact observed wavelength and the physical source of an emission line. This is because the emission line is assumed to be monochromatic and is dispersed in a specific direction and at a specific distance from the source. Figure \ref{fig:EM2Db} illustrates this, and shows how observations taken at multiple PAs of a resolved galaxy containing a star-forming region producing an emission line are related. The EM2D method inverts this approach: starting with emission line candidates identified directly in dispersed 2D images, the dispersion solution of the grism is inverted to determine where the source of the emission line might be located. For a single PA, the source of the emission line candidate can be anywhere along the spectral trace that goes through the detected emission line (shown as thick black lines in Figure \ref{fig:EM2Db}). When two PAs are available, the intersections of these multiple traces points to the actual source of the emission lines. Each time an intersect is found, it can be used to compute the observed wavelengths of the emission lines which, which, if real, should be identical, as shown in Figure \ref{fig:EM2Db}. In practice, the derived wavelengths are not identical, but they should be consistent when considering both position measurement errors of each emission line in the dispersed images and the accuracy of the existing grism calibration. This method automatically rejects zeroth spectra orders since these would result in widely inconsistent wavelength estimates when observed in different PAs. When more than two PAs are available, pairs of observations can be used to determine these intersects.  Multiple intersects at the same positions indicate more robust detections based on the emission line measured in independent observations with different PAs.  Thus, the more intersects (or the more PAs) are available, the more refined and the smaller the uncertainty/error in the computed observed wavelength. Figure \ref{fig:EM2D} demonstrates this with an example showing an object with two distinct star forming regions that would appear blended in the extracted spectra if using traditional slitless extraction methods.

\subsubsection{Detection of Emission Lines}\label{sec:EM2D_detection}
The first step of the reduction process is to use the reduced, background-subtracted, and astrometrically registered {\tt FLT} images and their associated Simulation-Based extraction (SBE), which are described in Section 3 of \cite{Pirzkal2017}. For each of the five available PAs of a given FIGS field, we used DrizzlePac's {\tt astrodrizzle} \citep{Avila2015} to produce a combined G102 dispersed mosaic and a combined simulated G102 mosaic by combining all of the observed data and simulated data, respectively. The simulated mosaic was then subtracted from the combined dispersed mosaic to produce a deep dispersed observation where the continuum light was removed. The initial emission line candidate list at each PA was generated by running SeXtractor \citep{Bertin1996} on the continuum-subtracted mosaics. All spatial features 2.5 $\sigma$\ above the local background were selected as candidate targets, yielding a few thousand candidates per mosaic. Next, the spectral trace was computed for each of the emission line candidates using the 5 available PAs. This was repeated for each of the four FIGS fields. The exact methodology and figures detailing the initial 2D extraction and resulting 1D spectra can be found in Sections 3.3.1 and 3.3.2 of \cite{Pirzkal2017}.

For all of the candidates, the intersect between two spectral traces was computed and the observed wavelengths of the  emission lines were obtained from these intersects.  If the inferred wavelength of the two line candidates differed by more than 48\AA\ (equivalent to 2 WFC3 pixels), the intersect was discarded. When using 2, 3, 4, 5 (and $n$) PAs, the maximum of possible pairs number is  1, 3 ,6, 10, and, generally speaking, $(n^2-n)/2$. A single region with a strong emission line could therefore appear as 10 separate intersects in a list. 
A robust list of emission line regions was selected using a density-based spatial clustering ({\tt DBSCAN}) algorithm \citep{Ester1996}. This algorithm groups together points closely packed together within a defined parameter space and rejects outliers in low-density regions of this space.  In this case, the R.A. and  decl.  and wavelength ($\lambda$) comprise the three-dimensional parameter space, and a cluster size of 0''.129 in RA and Dec and 25 \AA\ in wavelength was used to define intersects. The three parameters of each cluster were then averaged to produce the emission-line region candidate list. Each of these candidates has a grade equal to the number of intersects at this location and with this wavelength. Emission lines at different wavelengths, i.e. \OIIIg\ and \Hag\ were allowed to originate from distinct star forming regions, where any object can have more than one star forming region.  In fact,  no restrictions were imposed on the number or location of star forming regions, nor were they required to be located within known objects.  The advantage of this is that it permits for the discovery of naked emission lines, i.e. objects with clear emission lines that are not detected in the continuum.
%We note that no such emission line regions were detected (in contrast to a number of detections made in %PEARS).  All emission line regions in the FIGS fields were found to fall within the segmentation maps of %galaxies.

Although at least two intersects can improve the accuracy of both the wavelength of an emission line and its location within a galaxy, we required that an emission line be detected in at least 3 different PAs in order to be considered for further analysis.  The more stringent requirement assures that detections are robust against contamination, errant pixels, cosmic rays, etc. After this cut, a total of 1338 emission line candidates were identified in the four FIGS fields. Spectra of the corresponding star forming regions were extracted in a manner similar to that described in Sections 3.3.1 and 3.3.2 of \citet{Pirzkal2017}, but restricting the extraction width to  twice the size of the detected emission lines in the 2D dispersed images. Spectra from different PAs were combined together using a weighted average. Due to the very low S/N levels probed using the EM2D method, this initial list of emission line candidates required individual visual inspection to remove false positives, and to verify the quality of an emission line at the wavelength computed using the EM2D method. The first pass at line fitting allowed for fitting a single Gaussian profile as well as multiple Gaussian profiles to account for the \OIIg, \OIIIg, \Hb4861, \Hag, and \SIIg\ multiplets.  This final cut resulted in a final list of 338 emission line candidates within 302 emission separate emission-line regions (82, 53, 83, 83 in the GN1, GN2, GS1, and GS2 fields, respectively). This final list was then compared and correlated with the FIGS object catalogs \citep[Section 3.2.2 in][]{Pirzkal2017} in order to locate the host galaxies. These emission line candidates were produced by a total of 234 distinct objects (58, 47, 63, 66 in the GN1,GN2,GS1 and GS2 fields, respectively).

% N>=3
% GN1 248
% GN2 141
% GS1 373
% GS2 577

% from emmgradeV3.db
% grade values:
% 1256 in total
% 298 grade=1
% 90 grade=2
% 388 grade>0
% 274 IDs
% 94 GN1 grade>0
% 64 GN1 IDs
% 64 GN2 grade>0
% 56 GN2 IDs
% 117 GS1 grade>0
% 77 GS1 IDs
% 113 GS2 grade>0
% 86 GS2 IDs

The method presented here builds on the 2D emission line scheme for PEARS (PEARS-2D) first presented in \citet{Straughn2009} and \citet{Pirzkal2013}.  However, the main differences are that the PEARS-2D method required only a minimum of two PAs to accept an emission line detection, each with a peak at least 1.1$\sigma$ above the background. Emission  lines were detected in a continuum-subtracted image where the continuum was estimated by smoothing the data. The extraction relied purely on the reduction package {\tt aXe} \citep{Pirzkal2001,Kummel2009}.  The improved methodology and reduction used here, and described in more detail in \cite{Pirzkal2017}, significantly improve both background subtraction, and contamination by using SBE.

%[NOTE ON NAKED EMISSION LINES]

%\item Grism relation is used to compute the trace in ra,dec coordinates of the trace going through the detected lines (show fig)

\subsection{Emission-Line Identification}\label{EM2D:ID}
Once the spectra are extracted, the next step is to identify the detected emission lines in each of the 338 candidate objects.  The method of identification assumes the observed lines are \Lya, \OIIg, \Hb4861, \OIIIg, or \Hag\, as these are the brightest, most prominent rest-frame optical lines. Identification of emission lines is based on a classification scheme for the host galaxies.  Hosts are classified as Type I if the spectroscopic redshift of the host galaxy was already publicly available from sources in the literature. These redshifts were used as a starting point to identify the emission line and determine the final redshift of the source. Type II hosts are those for which we could identify multiple lines such as the \Hag\ + \SIIg\ or the \OIIIg\ + \Hb4861\ lines.  The \Hag\ and \SIIg\ lines are resolved in the G102 grism (see Figure \ref{fig:EM2D_fit}). Similarly, all three \OIIIg\ and \Hb4861 lines are resolved. Type II hosts  also include objects with multiple lines such as \OIIg\ together with \OIIIg, or \OIIIg\ together with \Hag. The G102 FIGS data were supplemented with existing archival G141 data (using data from proposals 11600, 12099, 12177, and 12461) extending wavelength coverage to 1.6 $\mu$m), which were processed and extracted in the same manner as the FIGS data. However, the G141 data were {\it only} used to detect other bright emission lines, such as spectroscopically confirming the identification of \OIIg\ in the G102 data.  We made limited use of the G141 data, because the G141 observations are significantly more shallow and suffer from heterogeneous coverage compared to the five-epoch G102 FIGS data.  This makes it problematic to compute the completeness levels for the G141 data.   Type III galaxies are those cases where neither publicly available redshifts, nor identification of multiple emission lines could be made. Instead, spectrophotometric redshift estimates of the host galaxy were used.  To compute these, photometric redshifts were first determined using the program BEAGLE \citep{Chevallard2016} and the available FIGS photometric catalogs. Next, a posterior distribution of the redshifts for each object was created to help identify an observed single emission line and compute a more accuate spectroscopic redshift for the source \citep[See,][]{Xia2011,Xu2007}.

\begin{figure}[h!]
    \centering
    \includegraphics[width=6in]{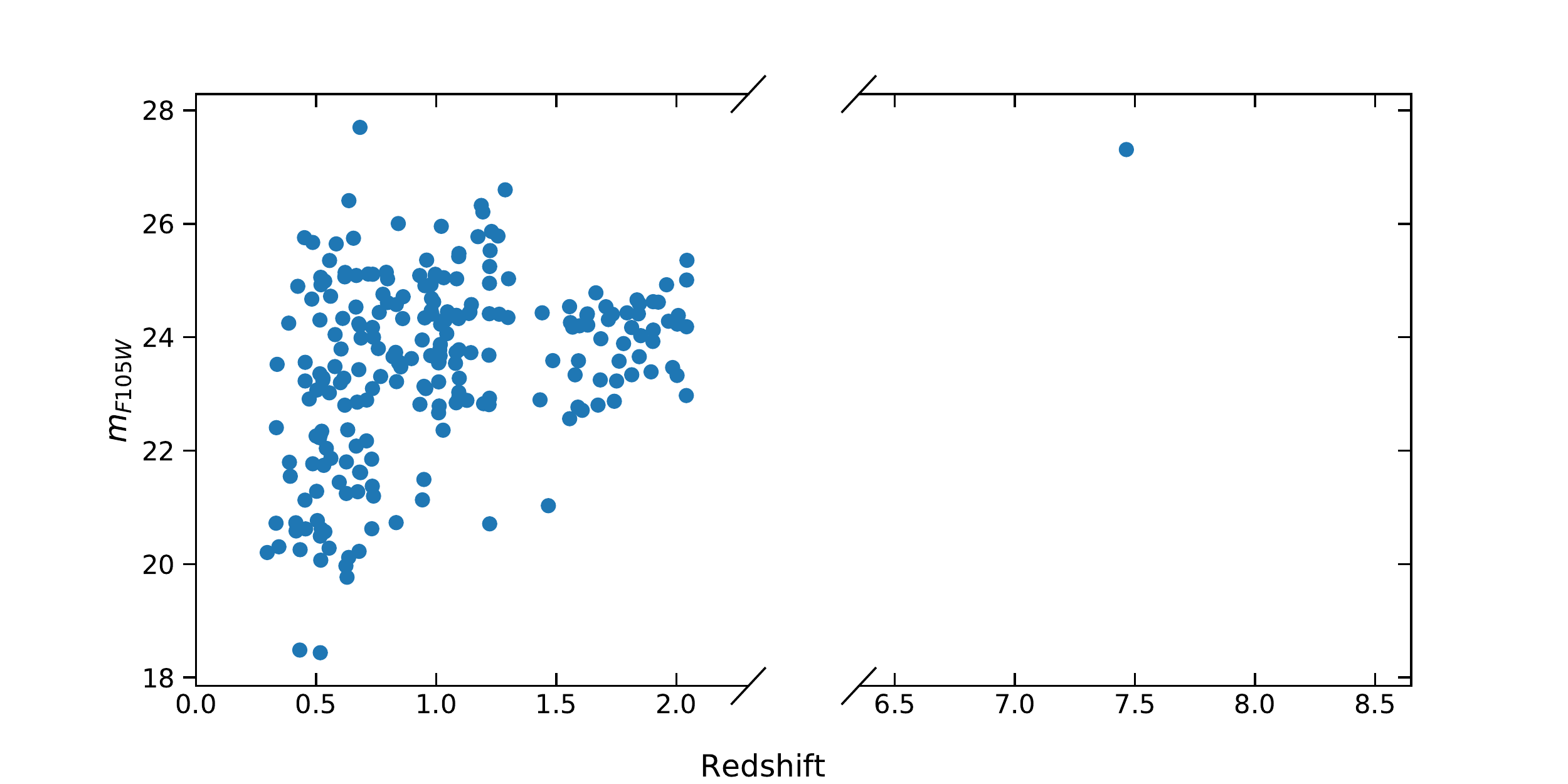}
    \caption{Host galaxies' apparent magnitude (F105W) as a function of redshift in the FIGS EM2D sample.}
    \label{fig:z_f105w}
\end{figure}

\begin{figure}[h!]
    \centering
    \includegraphics[width=6in]{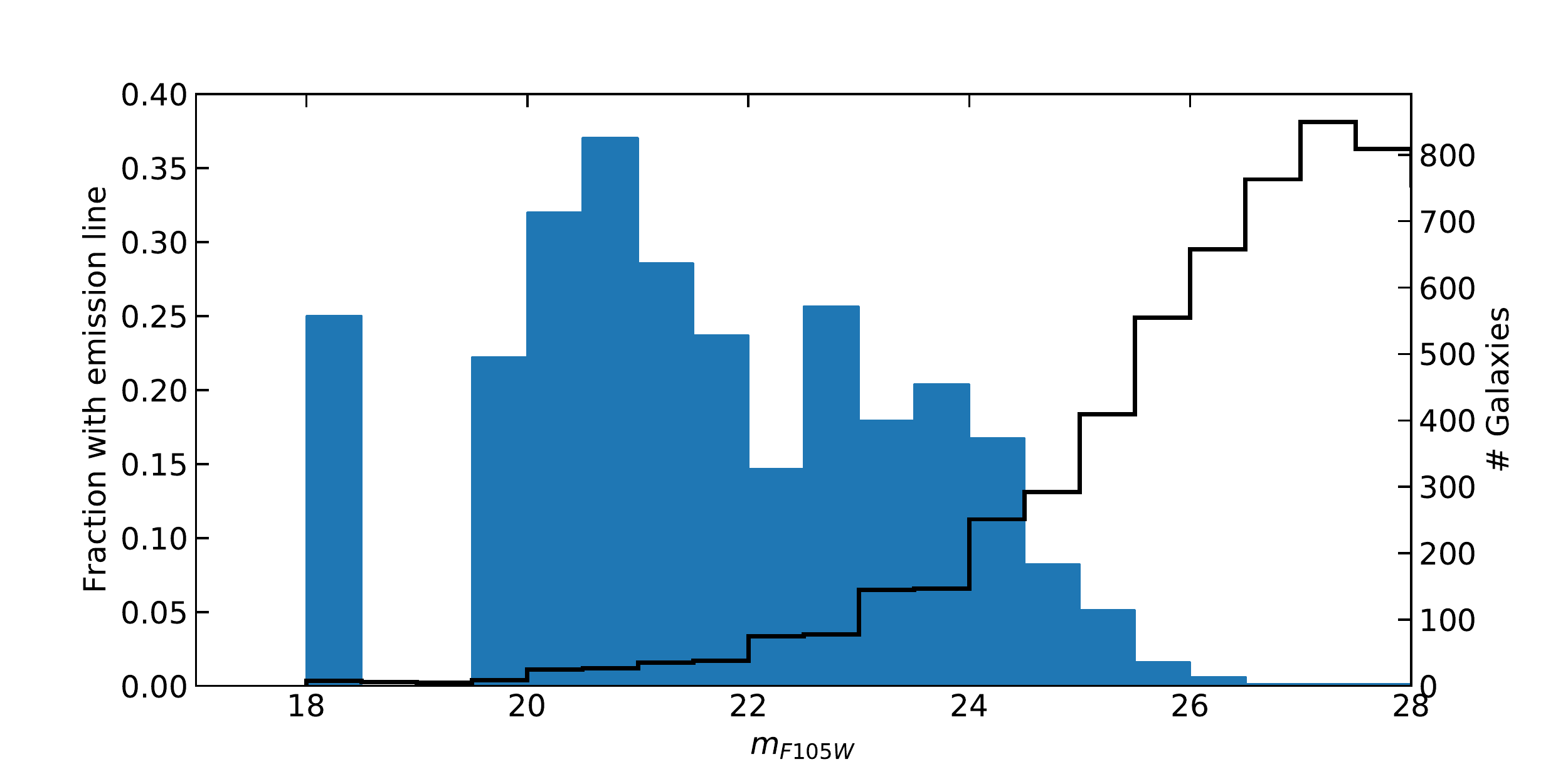}
    \caption{Fraction of FIGS galaxies observed in more than two position angles with an emission line region identified using the EM2D method (filled histogram). We also show the distribution of FIGS galaxies observed in more than two position angles. }
    \label{fig:emm_ratio}
\end{figure}

In some cases, candidate objects could be identified using more than one approach.  The order of priority for emission line identification is:  (1) ground-based spectroscopy; (2) grism spectroscopy; and (3) photometric redshifts (ground- and/or space-based).  When a spectroscopic redshift was in good agreement with our grism observations, we designated the host galaxy as Type I. When multiple emission lines in the slitless grism data improved upon a known spectroscopic redshift, we designated the host galaxy as Type II. In most cases, multiple emission lines were identified in a spectrum.  In a few rare circumstances, emission line identifications were uncertain. This occurred specifically when a single emission line was detected with the G102 slitless grism data, and was inconsistent with the photometric redshift of the host galaxy for all of the lines summarized above.  In these instances, such objects were removed from the sample.  The final sample then contains a total of 302 star forming regions in 234 distinct galaxies, 159 of which were spectroscopically confirmed as Type I or a Type II. A more detailed breakdown of our sample is shown in Tables \ref{tab:EMD2} and  \ref{tab:lines}, while Figure \ref{fig:z_f105w} shows the distribution of the host galaxy apparent magnitudes as a function of redshift. As we show in Figure \ref{fig:emm_ratio} we identified emission line regions in approximately 20\% of the entire FIGS sample, down to a continuum magnitude of {\it m}$_{F105W}=25$.

%Our final sample include 103 Type 1 sources, 105 Type 2 sources, and 102 Type 2 sources.

% 4 in ztype=0
% 103 in ztype=1
% 105 in ztype=2
% 102 in ztype=3
% 0 in ztype=4
\begin{deluxetable}{cccc} 
%\tabletypesize{\footnotesize} 
\tablewidth{0pt} 
\tablecaption{FIGS EM2D Emission-line Regions and Galaxies \label{tab:EMD2}} 
\tablehead{ 
\colhead{Field} & \colhead{Type} & \colhead{Number of Galaxies} & \colhead{Number of SF Regions}} 
\startdata 
GN1 &   I & 31 &    41 \\
GN1 &   II &    15 &    29\\
GN1 &   III &   12 &    12\\
\hline
GN1 &   Total &   58 &  82\\
\hline
GN2 &   I & 16 &    19\\
GN2 &   II &    14 &    16\\
GN2 &   III &   17 &    18\\
\hline
GN2 &   Total&    47 &  53\\
\hline
GS1 &   I & 23 &    34\\
GS1 &   II &    22 &    28\\
GS1 &   III &   18 &    21\\
\hline
GS1 &   Total&    63 &  83\\
\hline
GS2 &   I & 19 &    23\\
GS2 &   II &    19 &    24\\
GS2 &   III &   28 &    37\\
\hline
GS2 &   Total&    66 &  84\\
\hline
All &   I & 89 &    117\\
All &   II &    70 &    97\\
All &   III &   75 &    88\\
\hline
All &  Total& 234 & 302\\
\enddata 
\tablecomments{Breakdown of the number of distinct host galaxies and distinct emission-line regions detected and identified using the EM2D method. We also show the breakdown for Type I (spectroscopic), Type II (grism spectroscopic), and Type III (spectrophotometric redshift) line and therefore redshift identification. }
\end{deluxetable}

\begin{deluxetable}{cccccc} 
%\tabletypesize{\footnotesize} 
\tablewidth{0pt} 
\tablecaption{FIGS EM2D Emission Lines \label{tab:lines}} 
\tablehead{ 
\colhead{Line} & \colhead{GN1} & \colhead{GN2} & \colhead{GS1} & \colhead{GS2} & \colhead{Total}} 
\startdata 
\Hag + \SIIg & 44 & 11 & 39 & 44 & 138\\
\OIIIg + \Hb4861& 31 & 25 & 38 & 22 & 116\\
\OIIg & 7  & 16 & 10 & 26 & 59 \\
\Lya & 0 & 0 & 0 & 1 & 1 \\
\enddata 
\tablecomments{Number of individual \Hag, \SIIg, \OIIIg + \Hb4861, and \OIIg\ lines measured in the EM2D FIGS galaxy sample.} 
\end{deluxetable}

% EM2D Paper I Figures V9.ipynb

% EM2D Paper I Figures V9.ipynb

\subsubsection{Redshift Accuracy}\label{EM2D:z}

As noted in Section \ref{sec:EM2D_basic}, the de facto standard of wavelength identification is to assume that the source of the emission line is the broadband center of the object.  However, in cases where the emission line region and/or underlying host galaxy is resolved (or partially resolved), there will exist an offset between the actual source of the emission line and the broadband center of the object.  This offset results in an error in the value of the observed wavelength of the emission line. This error is proportional to the 
distance between the emission line region and the galaxy center.  Because EM2D is robust enough to detect emission line regions within a host galaxy, it can be used to significantly improve the accuracy of redshifts attained with slitless grism spectroscopy, approaching the level of accuracy one can obtain from higher-resolution slit spectroscopy.

The limiting factor in the accuracy of the EM2D method is the ability to accurately measure the exact position of the emission lines in the continuum-subtracted slitless observations (See Section \ref{sec:EM2D_detection}). Using a traditional centroiding method, emission lines with S/N=100 can be determined down to $\sim$ 0.1 pixel accuracy.  The error in the location of the star forming regions in galaxies is equivalent to a native WFC3 IR pixel, or $0.^{''}129$. The EM2D method can discern emission line candidates that differ by at most 25\AA\, so that the expected error in our wavelength calibration is better than $|\delta z| / (1+z) \approx 0.002$. This accuracy can be improved by increasing the number of position angles used. 

\begin{figure}[h!]
    \centering
    \includegraphics[width=6.5in]{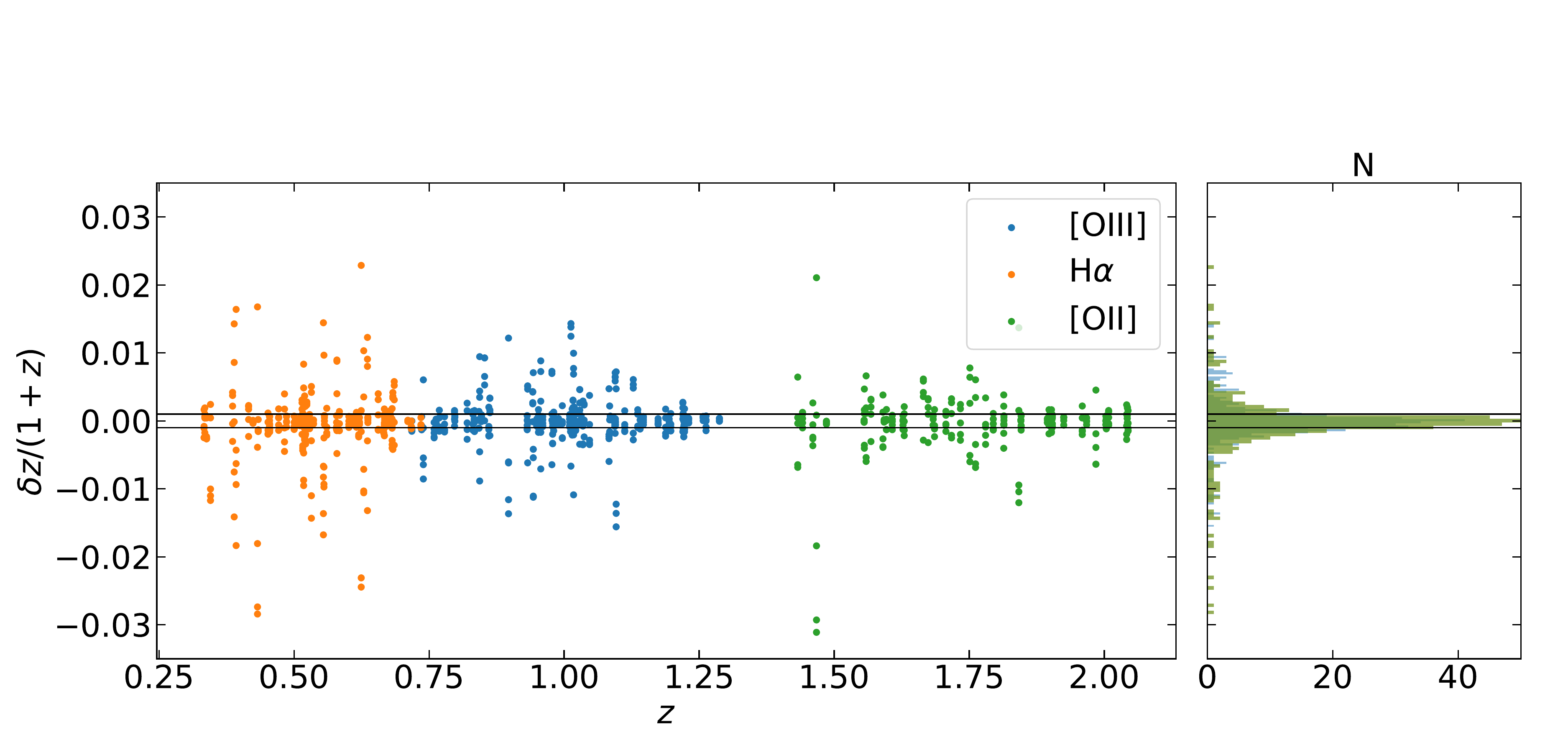}
    \caption{The error in the redshift estimates of the FIGS emission line galaxies ($\delta z / (1+z)$), if they had been observed using single PA observations, as a function of $z$ for the H$\alpha$, [OIII] and [OII] emission-line hosts. The solid black lines indicate $\delta z / (1+z) = \pm 0.001$. The panel on the right shows a histogram of the error in redshift.}
    \label{fig:EM2D_dz}
\end{figure}

We can empirically quantify the improvement made to the redshift determination of these sources using the FIGS data themselves, and calculating what observed wavelength we would have measured for every emission line in our sample when observed at each PA on the sky. In each case, the location of the emission lines in the host galaxy is used, and the resulting offset projected in the dispersion direction $\Delta x$.  The resulting error in wavelength is then $\Delta \lambda \approx \Delta x \times 25$\AA. This allows for the direct estimation of the effect of spatially offset stellar emission lines when using low-resolution slitless observations. For the purpose of this test, only objects with a single and robustly detected emission line region were used, namely those that have been detected in $n>=8$\ combinations of PAs (See Section \ref{EM2D:ID}).  Figure \ref{fig:EM2D_dz} demonstrates how large a fraction of the redshift derived for the FIGS sources would be off by  $|\delta z| / (1+z) > 0.01$, 
assuming that the source of the emission line is also the center of the source.  Taking into account that the resolution of the EM2D method is 1 native WFC3 G102 pixel, or $0.^{''}129$\ in the image and 25\AA\ in the spectra, 3\% of the redshift estimates of the FIGS sources would be off by  $|\delta z| / (1+z) > 0.01$, 8\%\  of the redshift estimates would be off by by $|\delta z| / (1+z) > 0.005$, and 12\%\ would be off by $|\delta z| / (1+z) > 0.001$. Thus, 2D methods such as EM2D can significantly improve the spectroscopic redshifts of ELGs by fully taking into account the location of the star forming regions within these galaxies.

% distRe_V9.ipynb

% \begin{figure}[h!]
%     \centering
%     \includegraphics[width=6in]{FIGS_dz}
%     \caption{The FIGS EM2D spectroscopic redshifts compared to known spectroscopic redshifts for source with secure spectroscopic redshift (Type 4 in Nimish's lists).}
%     \label{fig:FIGS_dz}
% \end{figure}

% \subsection{Implication on WFIRST etc}
% Knowing offset between the actual source an the emission line and the center of the object (which is used as wavelength calibration zero point) allows for a more accurate determination of the redshift of the source when using slitless spectroscopy.
% TBD: Redo/update the analysis in Ryan and Pirzkal 2016. Figure \ref{fig:EM2D_Rh}.

%\section{Line properties}

\subsection{Flux measurements}\label{sec:FIGS_MCMC}

\begin{figure}[h!]
    \centering
    \includegraphics[width=6in]{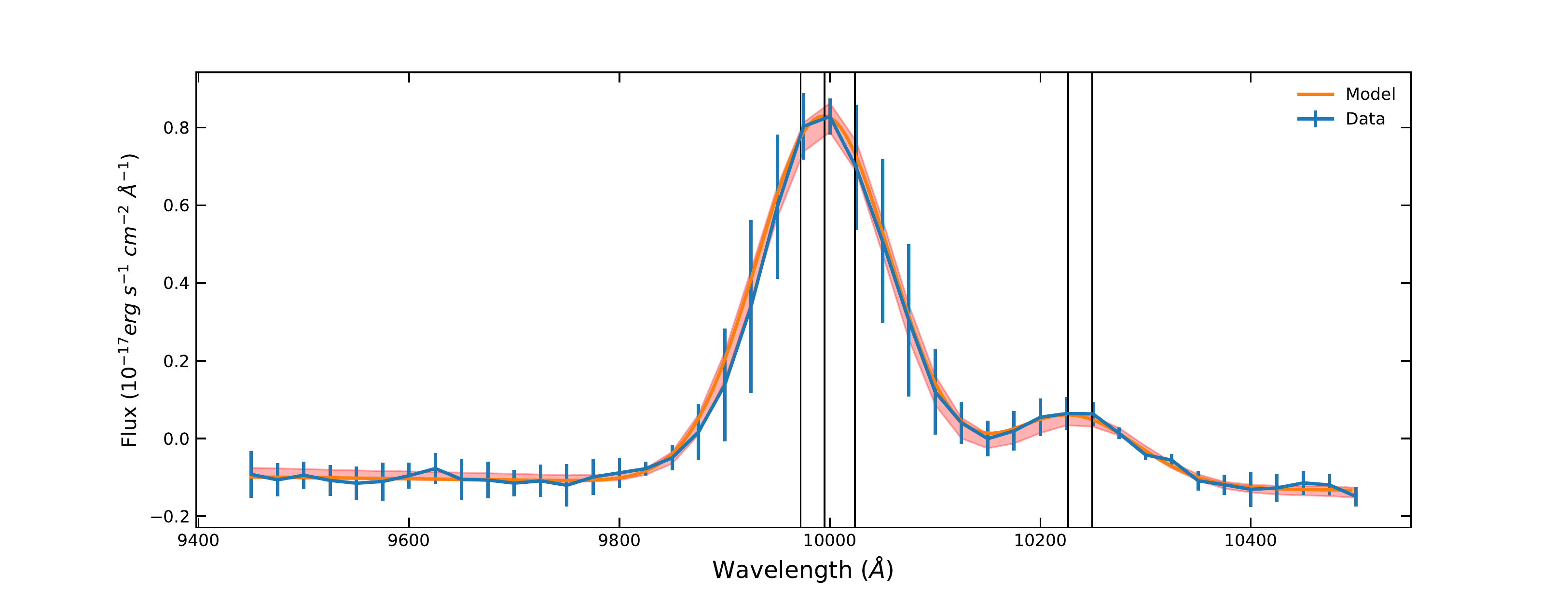}
    \caption{Example of an MCMC based fit of the \Hag\ and \SIIg\ lines. This is source GS2 3186. The blue line with error bar shows the data. The green line shows the formal best fit. Thin vertical lines show the location of the \Hag\ and \SIIg\ lines. As is often the case in slitless observations, the large line widths are caused by
the spatial structure of the source and the instrumental PSF and not kinematics.}
    \label{fig:EM2D_fit}
\end{figure}

\begin{figure}[h!]
    \centering
    \includegraphics[width=6in]{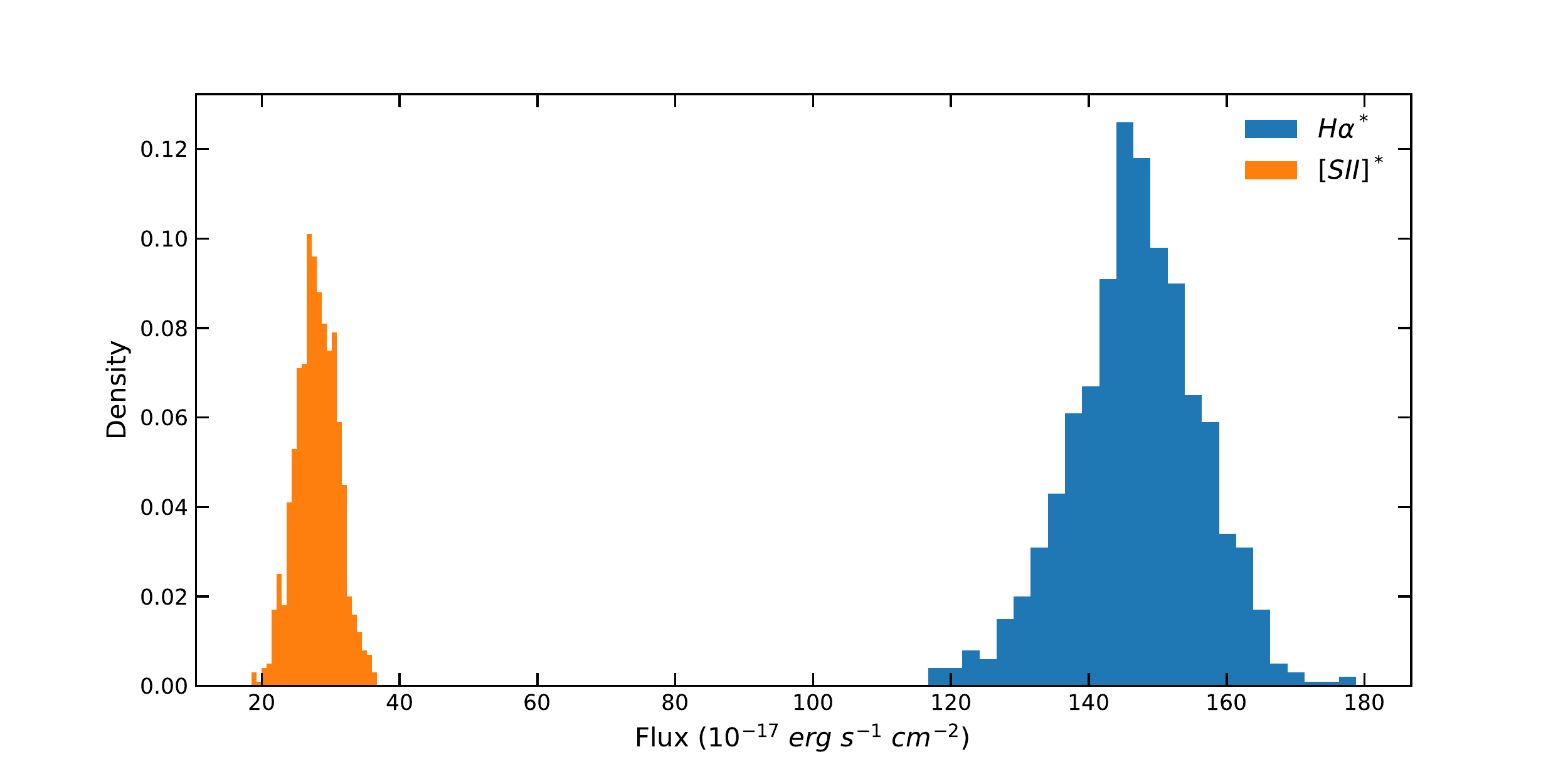}
    \caption{Histograms of the posterior distributions of the sum of the fluxes of the \Hag\ lines, and of the \SIIg\ lines in the spectrum shown in Figure \ref{fig:EM2D_fit}.}
    \label{fig:EM2D_pdf}
\end{figure}

\begin{figure}[h!]
\hbox{
    \centering
    \includegraphics[width=3in]{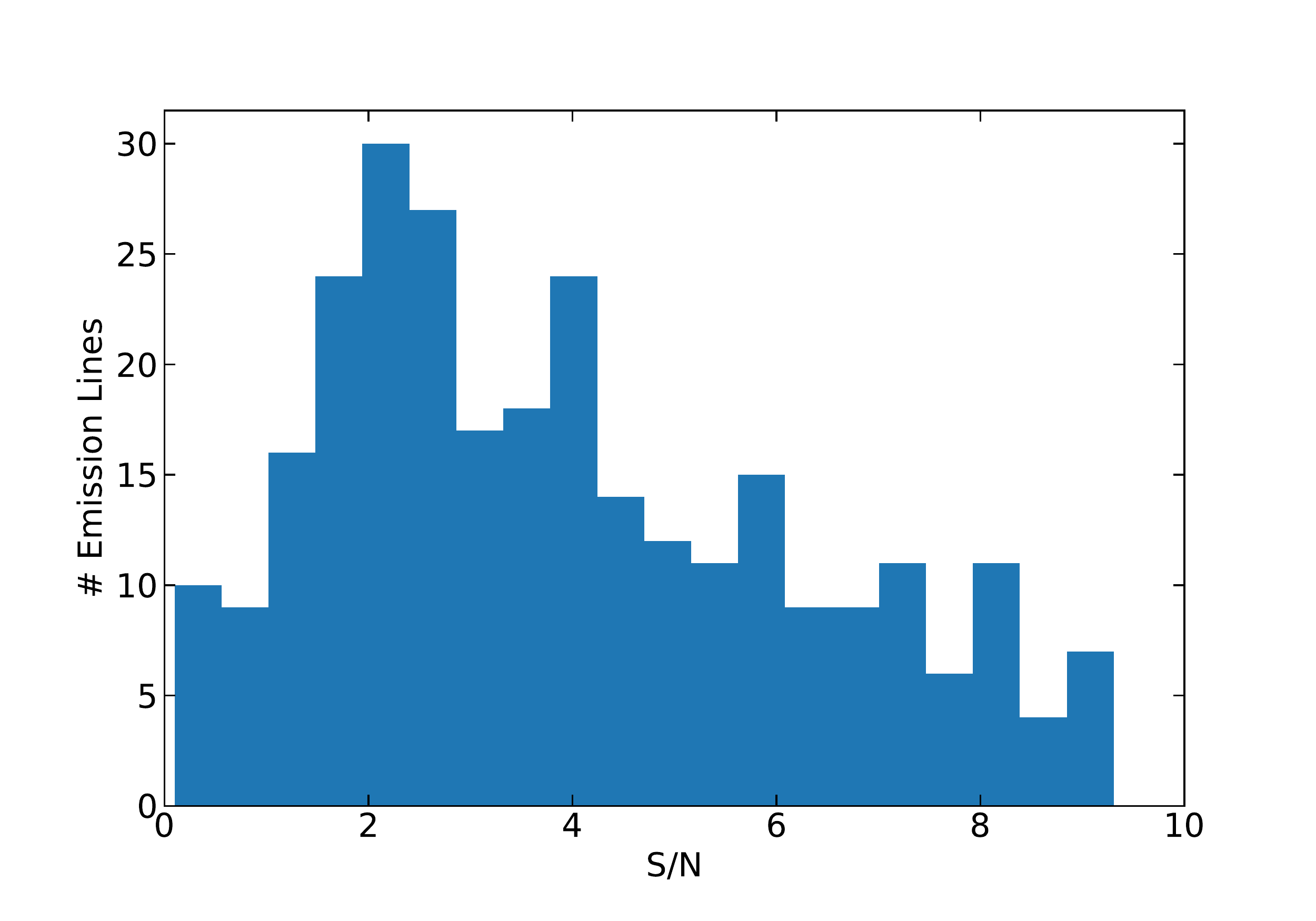}

    \includegraphics[width=3in]{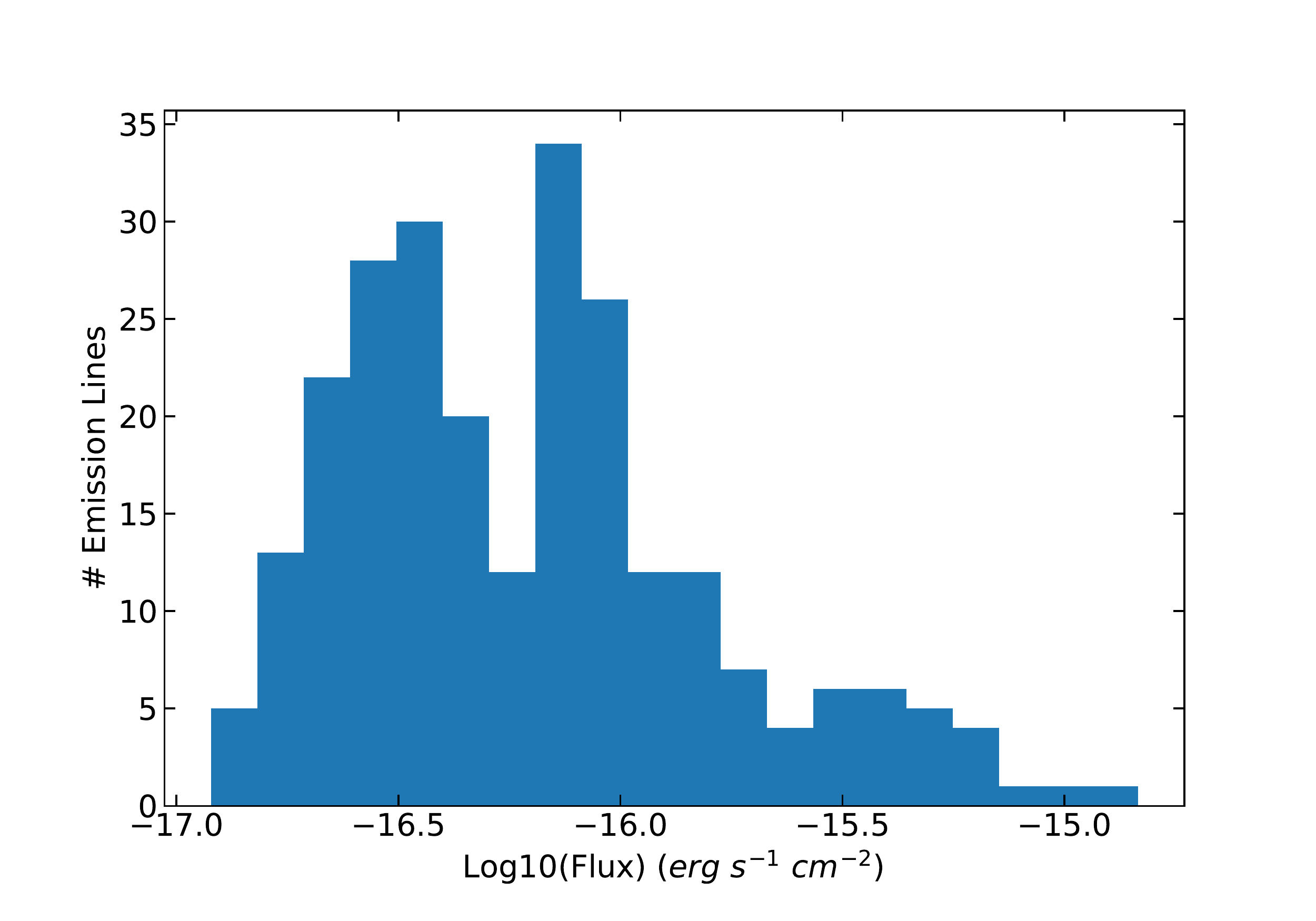}
}
    \caption{Left panel: distribution of the S/N of the measured line fluxes of \OIIg, \OIIIg, and \Hag\ in the emission-line regions identified in FIGS galaxies. Right panel: distributions of fluxes for emission lines with an S/N greater than 2.}
    \label{fig:EM2D_fluxes}
\end{figure}

The improved line identification and redshift constraints from EM2D were used to systematically fit the \Hag\ triplet, the \SIIg\ doublet, and the \OIIIg, the \Hb4861, and the \OIIg\ lines in the G102 spectra. A Gaussian profile was fit to each emission line, and a continuum level was estimated using the regions surrounding the expected positions of these emission lines. An initial fitting of each emission line was then followed by a Markov-Chain Monte Carlo (MCMC) based fitting method. This method is based upon the methodology and algorithms first described in \cite{Pirzkal2013}.  The MCMC fitting was used to define proper confidence intervals for the redshift of the source, the width of the emission lines (assumed to be the same for all the lines in a given spectrum), and the line fluxes. In the case of \OIIg\ the line flux ratio was defined to be within a range of $0.25<\lambda3726/\lambda3729<1.45$, following \citet{Osterbrock2006}. The \OIII5007, \OIII4959, and the \Hb4861\ lines were fitted using three gaussian profiles  with a flux ratio of \OIII5007\ to \OIII4959\ = 2.984, following \citet{Storey2000}. Finally, the \Hag\ and the \SIIg\ lines were fitted as three Gaussians with $0.4<\SII6717/\SII6731<1.42$, $\NII6583/\NII6548=3$ and $0.01<\Ha6464/\NII6583<1$, following \citet{Osterbrock2006}.  Figure \ref{fig:EM2D_fit}\  shows the best fit of the blended \Hag\ lines and the \SIIg\ doublet. The posterior probability distributions of the fluxes of the \Hag\ lines and the \SIIg\ doublet are shown in Figure \ref{fig:EM2D_pdf}.
The distribution of line fluxes that we measured is shown in Figure \ref{fig:EM2D_fluxes}, where we show the distributions of measured S/N and line flux of the FIGS EM2D emission lines.

The low resolution of the G102 grism does not allow for the lines to be fully resolved separately.  Therefore, the line fluxes we derive for the ${\rm H\alpha\ \lambda6563\ and\ the\ [NII]\lambda\lambda6548,6584}$\ lines are highly degenerate, and so we report the combined \Hag\ line fluxes. This is also the case for the the \SIIg\ lines. Similarly, the \Hb4861\ and \OIIIg\ were fit simultaneously, and in this case the wider separation of these lines allowed us to estimate the flux of these three lines independently. Finally, the \OIIg\ lines were fitted using two Gaussian profiles (although it is the sum of the flux from the doublet that is relevant for physical quantities such as the star formation rates) but as it is the case for the ${\rm H\alpha\ \lambda6563\ and\ the\ [NII]\lambda\lambda6548,6584}$\  lines, the derived line fluxes are highly degenerate.

% from FIT G102 G141 V9 MCMC.ipynb manually generated
% fitting Ha of GS1 1103, GS1_G102_1103_53.17396540_-27.77206169_Ha.pdf

% EM2D Paper I Figures V9.ipynb

\subsection{Locations of Emission Line Regions}\label{EM2D:loc}
\begin{figure}[h!]
    \centering
    \includegraphics[width=5.5in]{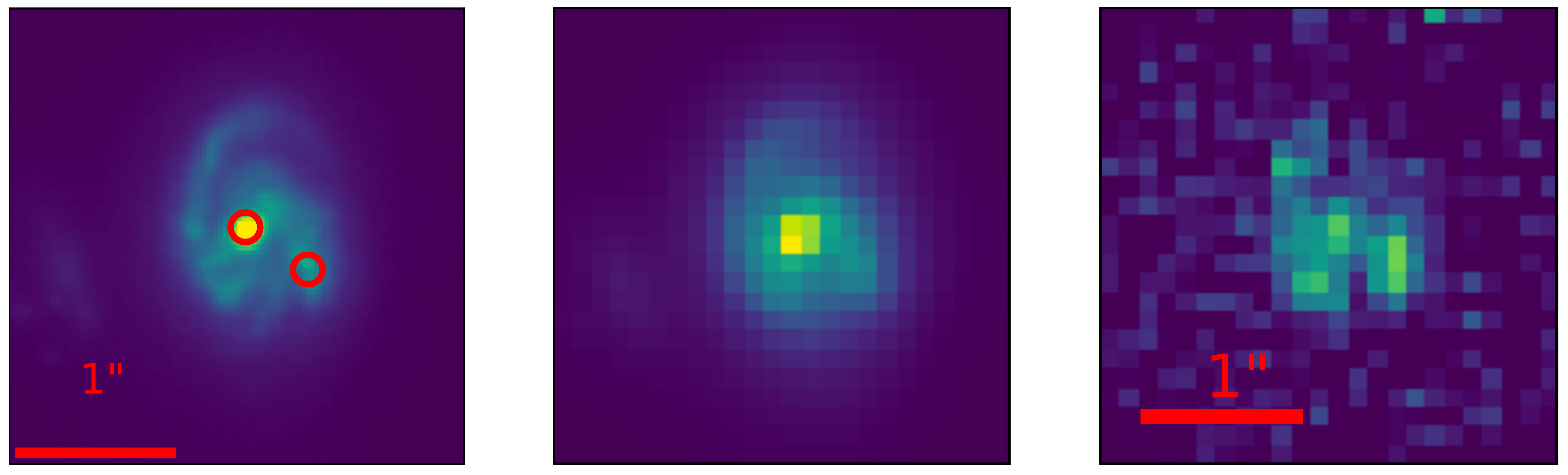}
    \caption{Left panel: galaxy GS1 1295 at $z=0.42$\ and the two \Hag\ emission-line regions (red circles) that we identified in this object using the EM2D method, as seen in the ACS F850LP filter. Middle pane: The same galaxy as seen in the WFC3 F105W IR filter.
    Right panel: a full 2D forward modeling of the \Hag\ emission in this object, which is facilitated by the accurate EM2D estimate of the observed wavelength of the emission feature. There is significant structure and extended line emission in this objects, and the two brightest ones were detected using the EM2D method. The other knots, seen in the right panel, are too faint to be detected by EM2D.
    }
    \label{fig:GS11295_All}
\end{figure}

As noted above, the EM2D method allows us to identify the position line emission within specific galaxies. In \citet{Pirzkal2013}, we showed that most emission lines are generated at a significant distance from the center of the galaxy over the redshift range of $0<z<1.5$. Using the near-IR FIGS data, we are now able to extend this work to the redshift range of $0.3<z<2.5$. Figure \ref{fig:GS11295_All} shows the star forming regions identified in a $z=0.42$\ galaxy.

% EM2D Paper I  Figures V9.ipynb

% EM2D Paper I  Figures V9.ipynb
\begin{figure}[h!]
    \centering
    \includegraphics[width=5.5in]{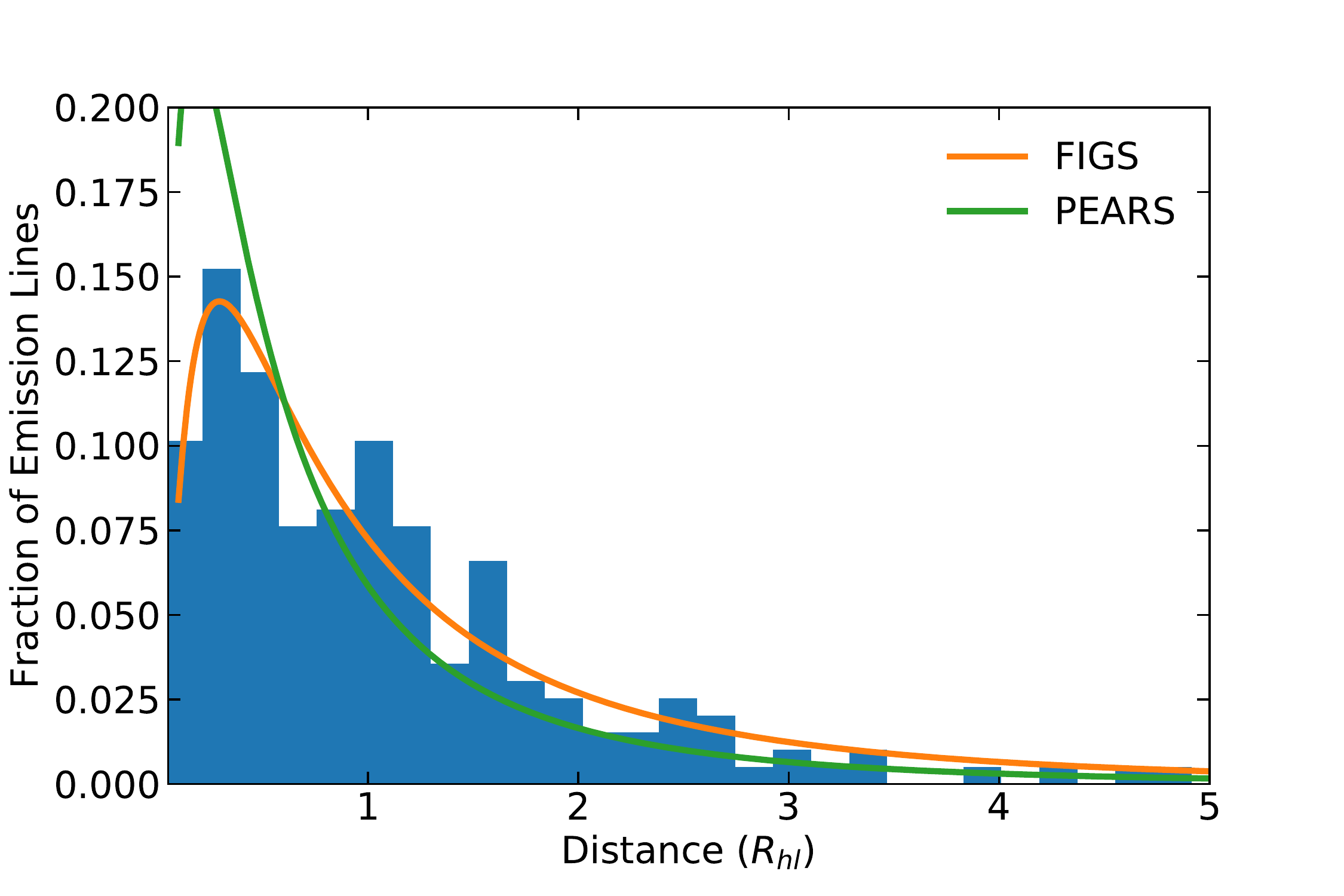}
    \caption{Distance of the FIGS emission line regions from the centroid of the host galaxies in units of galaxy half-light radius ($R_{hl}$). We show an exponential fit to the FIGS histogram in orange. A fit of the PEARS distribution from \citet{Pirzkal2013} is shown in green.}
    \label{fig:EM2D_Rh}
\end{figure}

In Figure \ref{fig:EM2D_Rh} we plot the distribution of emission line positions, normalized to the half-light radius ($R_{hl}$) of the host galaxies. The latter was computed by measuring the $R_{hl}$ of the galaxies in all available broadband HST imaging, and by computing the rest-frame $R_{hl}$ by interpolating the values measured in different bands (ACS F435W, F606W, F775W, F814W, F850LP, and WFC3 F105W, F125W, F140W, and F160W where available). As we show in this Figure, the distribution peaks at $\approx 0.3 R_{hl}$ and $\approx 44\%$\ of emission lines originate more than one $R_{hl}$\ away from galaxy center, where the peak of the distribution is a consequence of the limited spatial resolution of the FIGS survey. This observed distribution is consistent with the one we derived in \citet{Pirzkal2013}, although with a slightly larger proportion of knots being at larger distances. Exponential fits to the PEARS and FIGS distributions are also shown in Figure \ref{fig:EM2D_Rh}.

\subsection{Extended/diffuse emission}\label{sec:EM2D_ext}

We note that a majority of the galaxies in the FIGS sample have unresolved, or only marginally resolved, star forming regions. A small number of objects show signs of physically extended and/or diffuse emission. The extended sizes of these emission-line regions result in the apparent appearance of broadened lines in the extracted FIGS spectra. With the resolution of the G102 grism of $\approx 36$\AA, our ability to measure velocity dispersion is thus limited to about ${\rm 1000\ km\ s^{-1}}$, which is much greater than what one would expect from the dark matter halos in these galaxies. Some extended emission line regions were first flagged during the manual line classification process described in Section \ref{sec:EM2D_detection}. Here, we found 53 extended \Hag\ regions (in 41 distinct sources), and 9 extended \OIIg\ regions. Defining a region to be spatially extended if the measured emission lines FWHM are wider than 50\AA, we find that approximately 45\% of the \Hag\ regions are extended, while only 17\%\ of the \OIII5007 and \OIIg\ regions are spatially extended.
We can expect some of these objects to have extended active galactic nucleus (AGN) emission lines. In order to estimate the fraction of the FIGS EM2D sources we expect to be caused by an AGN, we cross-correlated our emission line catalog with the X-ray catalogs of \citet{Luo2017, Xue2016, Villforth2010}. Assuming a 1 arsec matching radius between the two, we find that $\approx4\%$\ of the FIGS emission lines are potentially AGN-driven. This fraction increases to $10\%$\ if we allow for a 5 arcsec matching radius. Of the 41 objects showing spectroscopically resolved \Hag\ emission, we find that 3 (7) sources are detected in the X-ray when using a matching radius of 1 (5) arcsecond. Those AGN-dominated candidates are GN1 1497, GS1 2614, and GS2 1653 using a 1 arcsec matching radius, and GN1 1497, GS1 2614, GS1 2363, GS1 2518, GS1 4308, GS1 1299, and GS2 1653, using a 5 arcsec matching radius. We therefore find 7\% (17\%) of our extended \Hag\ sources to possibly be AGN-dominated.

In the cases of larger galaxies, extended emission can lead to multiple detections using the EM2D method, and can cause artificially high $n$\ values to be assigned to a single star formation knot as several knots are found to be very close together, and merged during the clustering step described above. We found 15 sources that show signs of extended emission based on this criterion ($n>10$) alone. The galaxy shown in Figure \ref{fig:GS11295_All} is one of these. Once the observed wavelength of these emission lines is accurately derived, one can easily use this information to forward model the continuum-subtracted 2D spectra and reconstruct full 2D emission line maps. The result of the \Hag\ 2D map reconstruction for the same galaxy is shown in Figure \ref{fig:GS11295_All}. Our implementation of this 2D reconstruction will be fully described in the next paper in this series, where we describe the physical properties of emission-line selected FIGS galaxies in details.

% \begin{figure}[h!]
%     \centering
%     \includegraphics{GS1_1295_2D}
%     \caption{The same object shown in Figure \ref{fig:GS11295} and where we detected two strong emission line knots using EM2D, allowing us to derive a secure redshift of 0.42. The left panel shows the WFC3 F105W IR image of this source. The right panel shows a full 2D forward modeling of the \Hag\ emission in this object, which is facilitated by the accurate EM2D estimate of the observed wavelength of the emission feature. There is significant structure and extended line emission in this objects, and the two brightest ones were detected using the EM2D method, as we showed in Figure \ref{fig:GS11295}. }
%     \label{fig:GS11295_2D}
% \end{figure}

\subsection{Additional Emission Lines}
The flux of additional emission lines was also measured. For each spectrum, we fitted Gaussian profiles to additional emission lines, assuming that these were at the observed wavelengths determined using the redshifts determined during the fit of the \Hag\ triplet, the \SIIg\ doublet, the \OIIIg, the \Hb4861, and the \OIIg\ lines. We used the same MCMC based flux measurement technique that we described above while fixing the width of the emission lines to the value computed for the brighter emission lines. We performed flux measurements for the HeI $\lambda$5877, MgII $\lambda$2799, OI $\lambda$7774, [ArIII]$\lambda\lambda$7136, [ArIII]$\lambda\lambda$7753 lines as well as the 
H$\gamma$ $\lambda$4342/[OIII]$\lambda$4363, [NeIII]$\lambda\lambda$3869/[NeIII]$\lambda\lambda$3890, [NeV]$\lambda\lambda$3347/[NeV]$\lambda\lambda$3427, [OI]$\lambda\lambda$6302/[SIII]$\lambda\lambda$6312 potentially blended lines. We estimate that out of a possible 770 additional line detections, at most 4\% of the FIGS spectra show tentative evidence of emission at these additional wavelengths ($>2\sigma$\ detection in objects that are not overly extended with ${\rm FWHM<40\AA}$). The measured line fluxes of these additional emission line are between $5 \times 10^{-18}\ erg\ s^{-1} \ cm^{-2}$\ and $2.5 \times 10^{-17}\ erg\ s^{-1} \ cm^{-2}$.

%Line most detected where [NeIII]$\lambda\lambda$3869 (6), HeI $\lambda$5877 (5), H$\gamma$ $\lambda$4342/[OIII]$\lambda$4363 (7), OI $\lambda$7774 (5).

\subsection{Completeness and survey area}\label{sec:completeness}
The EM2D method is a relatively complicated method, and it is important to quantify its ability to detect emission lines. We determined the completeness function separately for each of the four FIGS fields, and for the combined flux of the \Hag\ lines, the flux of the \OIII5007\ line, and for the combined flux of the \OIIg\ emission lines. Emission-line regions were added randomly within individual objects in the field, and used to create dispersed simulations of emission line regions with emission lines of varying observed fluxes and at varying observed wavelengths.
These simulated data were then processed using the EM2D method like the original FIGS data. These simulations showed that our ability to detect emission lines in the dispersed images was strongly dependent on the contrast ratio between the simulated line flux and the dispersed broadband flux of the host galaxy at the location of the emission-line region. We define the contrast ratio ${\cal C}={F(line)\over F(cont.)}$\ as the ratio between the flux produced in the emission line region, assumed to be from one strong emission line such as [OII], [OIII] or \Hag, and the local broadband flux of the host galaxy. 

\begin{figure}[!h]
    \centering
    \includegraphics[width=5in]{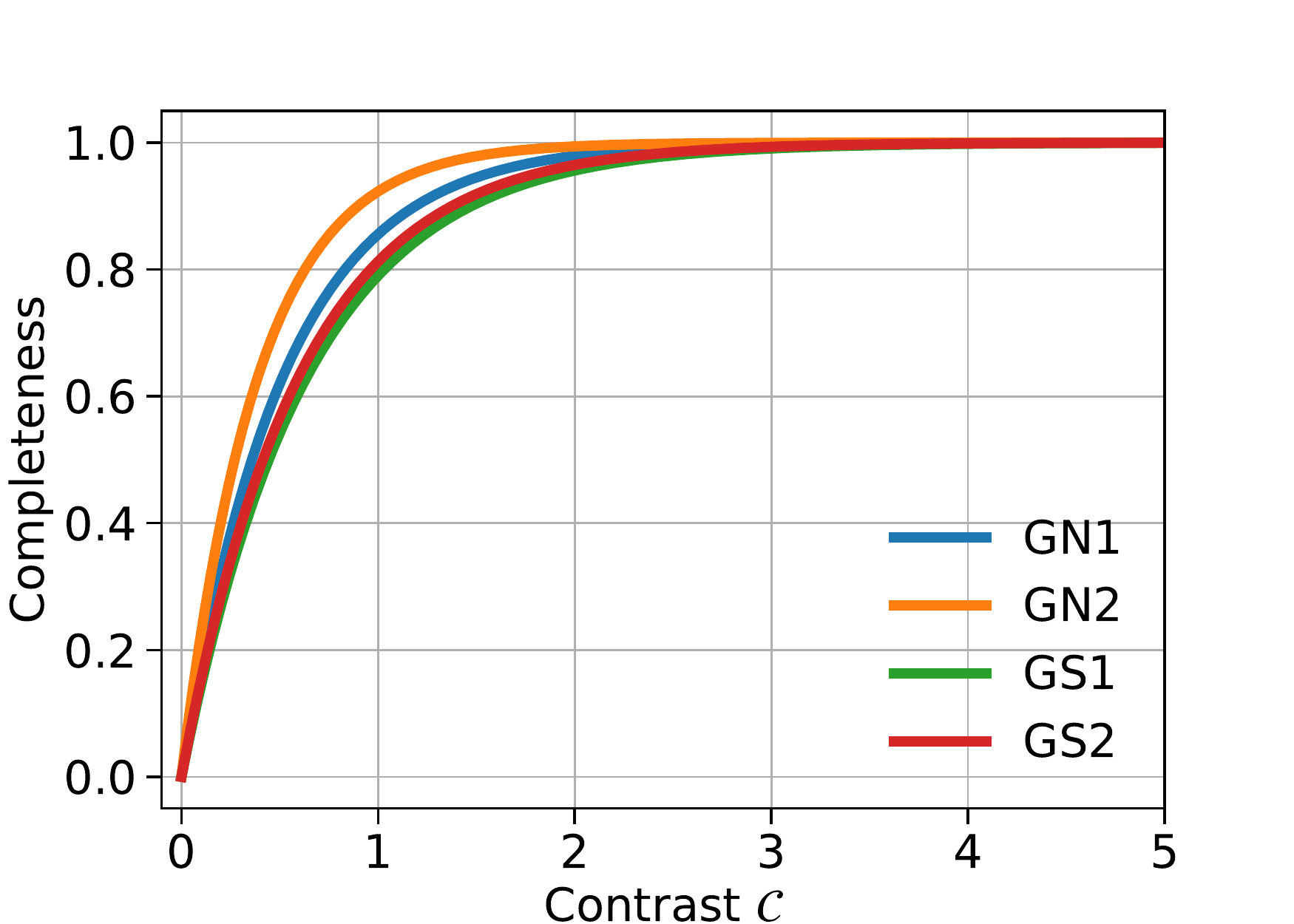}
    \caption{Completeness function f($\cal C$) for the EM2D FIGS emission lines as a function of ${\cal C} = {F(line) \over  F(cont.)}$.}
    \label{fig:FIGS_completeness}
\end{figure}

The fraction of emission lines that we recovered, using the same criteria as listed in Section \ref{sec:EM2D_detection}, was tabulated as a function of $\cal{C}$. This process was carried out 200 emission lines at a time, and a total of 40,000 emission lines were generated using this process. Figure \ref{fig:FIGS_completeness} shows a  plot of the completeness function for the FIGS survey as a function of $\cal{C}$.
We also used emission-line simulations to determine the effective area of each of the FIGS fields. Since the G102 grism observations are rotated and offset with respect to each other, the exact area of overlap is a function of the different PAs at which a FIGS field was observed. To quantify this process, we generated a sample of bright emission line, well within the wavelength range of the G102 grism, both in and outside each of each of the FIGS fields field of view. We then kept track of the locations on the sky from which emission lines were detected using the EM2D method using the same selection criteria used for the actual FIGS data. This process allowed us to estimate that the effective areas of the four FIGS grism fields are 4.66, 4.69, 4.62, and 4.21 arcmin$^2$\ for the GN1, GN2, GS1, and GS2 fields, respectively. Note that the smaller effective size of the GS2 field is a direct result of a very bright source being at the center of the field.

% \begin{figure}[!h]
%     \centering
%     \includegraphics[width=7in]{EM2D_Completness_v9}
%     \caption{Completeness function f(z,L) for the \Hag\ lines ($H_\alpha^*$), \OIII5007\ and \OIIg\ lines ($[OII]^*$) (left to right panels) emission lines in the FIGS EM2D emission line sample. We also show actually detected emission lines in red.}
%     \label{fig:FIGS_completeness}
% \end{figure}

\section{Results}\label{sec:results}

\subsection{''naked'' emission lines}\label{sec:results_naked}
\begin{figure}[h!]
    \centering
    \includegraphics[width=5.0in]{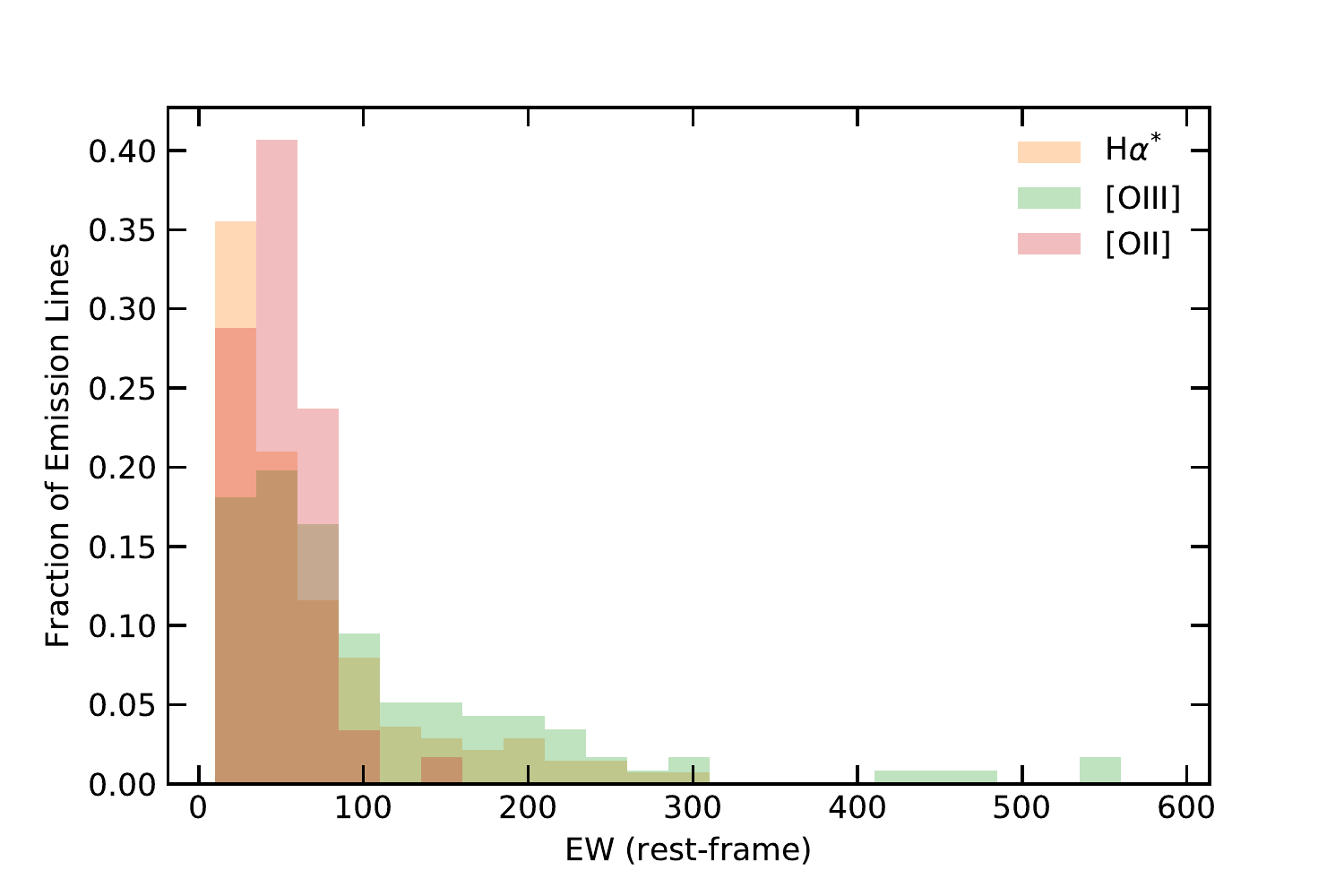}
    \caption{Distribution of rest-frame Equivalent Widths for the \OIIg\ ($[OII]^*$), \OIII5007, and \Hag\ ($H\alpha^*$) emission lines.}
    \label{fig:EM2D_EW}
\end{figure}

\begin{figure}[h!]
    \centering
    \includegraphics[width=3.0in]{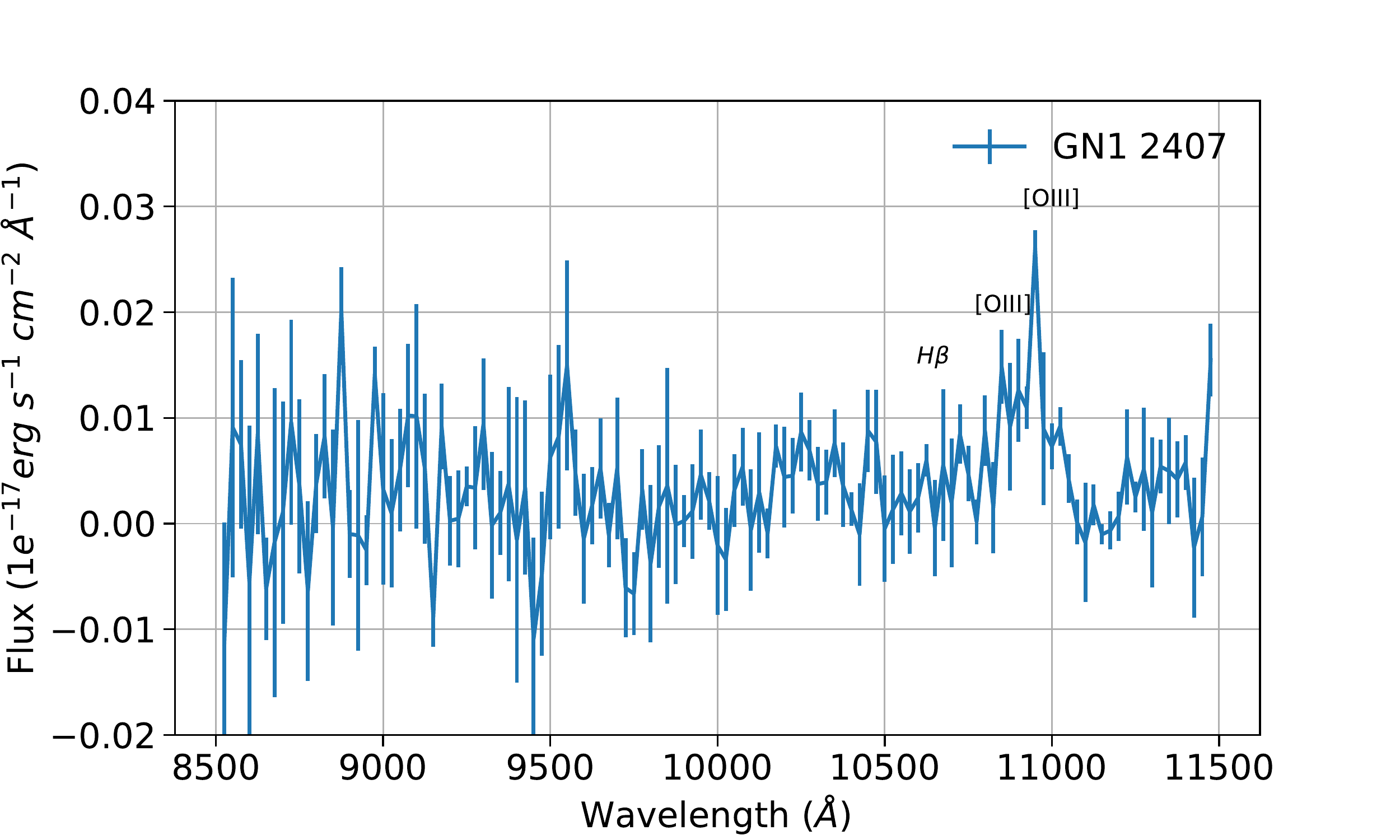}
    \includegraphics[width=3.0in]{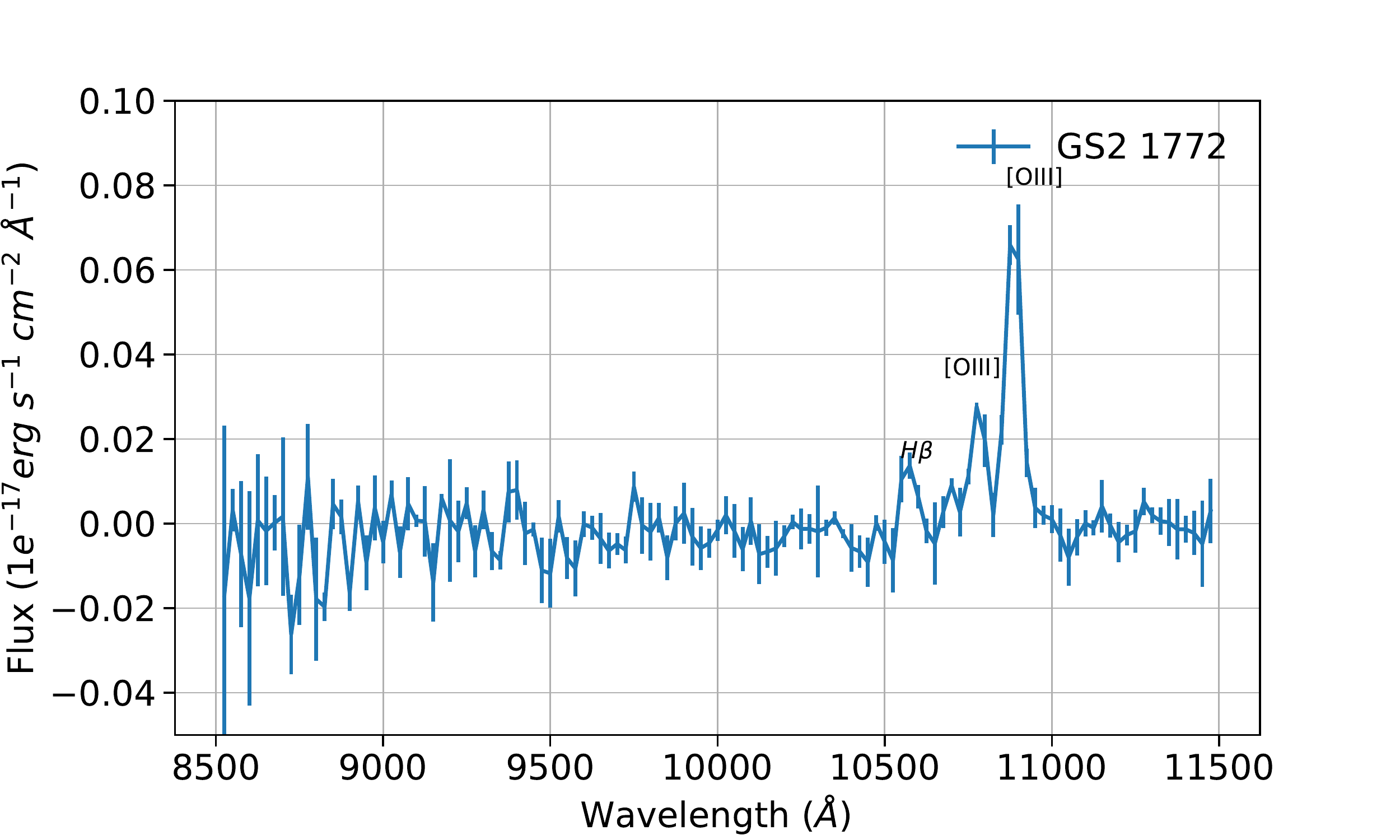}
    \caption{Spectra of two faint galaxies with large EW [OIII] lines. The host galaxies GN1 2407 and GS2 1772 have measured broadband F105W magnitudes of $m_{AB}=27.29$\ and $m_{AB}=26.25$} and line fluxes of ${\rm 1.46 \pm 0.22 \times10^{-17}\ erg\ s^{-1}\ cm^{-2}}$ and ${\rm 4.34 \pm 0.32 \times10^{-17} \ erg\ s^{-1}\ cm^{-2}}$, respectively. The measured rest-frame equivalent widths of the [OIII]\ 5007\AA\ lines are ${\rm 254\AA\pm63\AA}$\ and ${\rm 482\AA\pm69\AA}$, respectively.
    \label{fig:EM2D_naked}
\end{figure}

EM2D identifies emission lines independently of their host galaxies. We note that in every single case, we were able to easily identify an EM2D emission region with its host galaxy using nothing more than the SeXtractor segmentation map information \citep{Bertin1996}. We found no occurrence of a confirmed emission line region that was not formally within the segmentation map area of a galaxy. As we show in Figure \ref{fig:EM2D_EW}, we did, however, identify emission line regions with very large EWs and that are located within faint host galaxies. Examples of such sources are the two galaxies GN1 2407 and GS2 1772 shown in Figure \ref{fig:EM2D_naked}. These objects have bright [OIII] emission lines with observed fluxes of ${\rm 1.5\times10^{-17}\ erg\ s^{-1}\ cm^{-2}}$ and ${\rm 4.39\times10^{-17}\ erg\ s^{-1}\ cm^{-2}}$, respectively. The continuum sources, with $m_{AB}=27.29$\ and $m_{AB}=26.25$ in the WFC3 F105W filter, are estimated to be at the redshifts of $z=1.18$\ and $z=1.17$, respectively. We estimate these emission lines to have rest-frame equivalent widths (EWs) of 254\AA\ and 482\AA, respectively. At the lower redshifts of $0.112<z<0.36$\ and $z<0.05$, \citet{Cardamone2009} and \citet{Yang2017} have respectively identified strongly SFGs known as Blueberries and Green Peas. Based solely on an [OIII] EW that is larger than 300\AA, we identify 5 strongly star forming regions in the EM2D FIGS galaxies at $0.77<z<1.29$\ out of a sample of 116 [OIII] emission-line regions. 

In terms of emission line galaxies, we find 5 [OIII] galaxies with at least one ${\rm EW>300}$\AA\ star forming region out of 96 galaxies, 5\% of our sample. For comparison, we find that out of $\approx500,000$\ SFGs at $z<0.6$\ in SDSS data release DR10 with emission-line properties extracted by the Portsmouth group reported in the SDSS EmissionLinesPort table, 7\% of the galaxies have [OIII] line rest-frame ${\rm EW>300}$\AA.

% 2407 27.29 1.47731329138e-17 EW=556. [OIII] z = 1.18
% 1772 2625 4.38846545698 EW=1048.4131542 [OIII] z= 1.17

%TBD: Note on some super faint sources >27AB with prominent lines like GN1 2407 and GS 1772

\subsection{High Redshift Sources}\label{sec:results_highz}
The EM2D method successfully identified one \Lya\ emitter. This object (GS2 1406), while already known \citep{Larson2018} was independently detected using the EM2D method. The \Lya\ emission was blindly detected in 5 PAs, and associated with object GS2 1406. Based on the EM2D method and assuming that the emission line is \Lya, we estimate the redshift of this source to be 
$7.464^{+0.004} _{-0.006}$. We measure a \Lya\ line flux of ${\rm 1.58 ^{+0.6} _{-0.2} \times 10^{-17} erg\ cm^{-2}\ s^{-1}}$, and a very high equivalent width of  $172\AA\pm46\AA$, which is consistent with the results shown in \citet{Larson2018}. A second known high redshift source at $z=7.51$\ \citep[GN1 1292,][]{Tilvi2016} only shows emission lines in two position angles and is therefore not included in our FIGS EM2D sample, which required an emission line to be detected in at least three PAs (See Section \ref{sec:EM2D_detection}).

\subsection{Line Luminosities and Luminosity Functions}\label{sec:lum}
\begin{figure}[!h]
    \centering
    \includegraphics[width=5in]{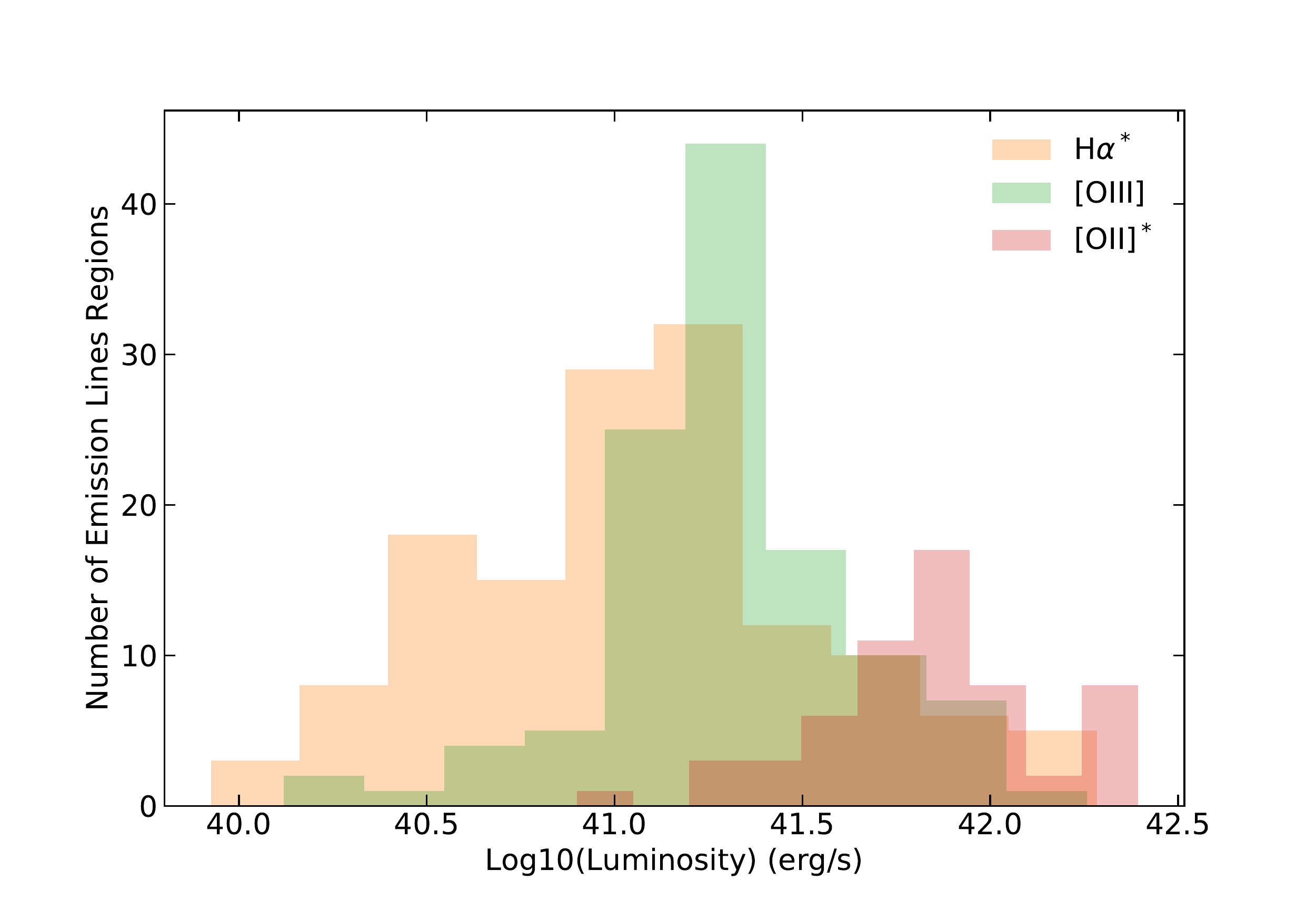}
    \caption{Distribution of the luminosities of individual emission line regions. The luminosities we show are  for the \OIIg\ doublet ([OII]$^*$), the sum of the \Hag\ lines (H$\alpha^*$), and the \OIII5007 line.}
    \label{fig:EM2D_Luminosities}
\end{figure}

Figure \ref{fig:EM2D_Luminosities} shows the distributions of \Hag\ lines, \OIII5007\ and the \OIIg\ emission line luminosities.
Combining our line catalog and the completeness, we derive the luminosity function of the \Hag\, \OIII5007\ and \OIIg\ emitters. 
The luminosity function was determined using the $1/V_{max}$ method, which we can express as:

\begin{equation}
    {\rm \Phi(log L_i) = {1 \over {\Delta log L}} \sum_j {1 \over {V_j}}}
\end{equation}

where ${\rm \Delta log L}$\  is the logarithmic bin width of the luminosity function, and $V_j$\ is the maximum volume within which emission line $j$ at a redshift of $z_j$ would still be included in our sample. The volume $V_j$, when accounting for the completeness of our sample (Section \ref{sec:completeness}), can be computed as:

\begin{equation}
    V_j = {\Omega \over {4 \pi}} \int_{z_{j,min}}^{z_{j,max}} R(\lambda) f({\cal C}) {dV_c(z) \over dz} dz
\end{equation}
Here, $\Omega$\ is the solid angle of our survey (in sr), $V_c$\ is the comoving volume element at redshift $z$,  $f({\cal C})$\ is the completeness function that we described above, and $R(\lambda)$\ is the normalized sensitivity function of the G102 grism. The latter takes a value of zeros outside of the bandpass of the G102 grisms, so that the values of ${z_{j,min}}$ and ${z_{j,max}}$\ can therefore be taken to be 0 and $\infty$, respectively.

The distribution of observed, non-dust-corrected line region luminosities is shown in Figure \ref{fig:EM2D_Luminosities}. We performed a systematic search for 2D emission lines using ACS on HST as part of the previous program PEARS. The use of the G102 grism by the FIGS survey, however, extends the redshift ranges to an epoch that is closer to the peak of star formation with  redshift ranges of $0.3<z<0.72$, $0.70<z<1.26$, and $1.28<z<2.0$ for \Hag, \OIII5007, and \OIIg, respectively. 
The FIGS complements the PEARS survey and the two surveys overlap in the redshift ranges that they probe: $0.3<z<0.45$, $0.7<z<0.9$, and $1.28<z<1.54$\ for the \Hag, \OIII5007, and \OIIg\ lines, respectively. The number of available FIGS sources within these redshift ranges is small (10, 18, and 5, respectively), but we can still compare the average volume densities derived from both surveys. We find that the mean volume densities agree well, as we estimate these to be $0.0015 \pm 0.00034$, $0.00055 \pm 0.000055 $, $0.00012 \pm 0.000035$\ $Mpc^{-3}$\ for the FIGS survey versus $0.0016 \pm 0.00028$, $0.00035 \pm 0.000069$, $0.000099 \pm 0.000024$\ $Mpc^{-3}$\ for the PEARS surveys, for the \Hag, \OIII5007, and \OIIg\ lines, respectively.

% EM2D Paper I Figures V9

We can compare our H$\alpha$\ observations to previous ground-based studies of H$\alpha$\ emission lines such as the one from  \citet[PEARS $z=0.26$,][]{Pirzkal2013}, \citet[WySH $z=0.40$,][]{Dale2010}, \citet[DAWN $z=0.62$,][]{Gonzalez2018},
\citet[$z=0.84$,][]{Villar2008},
\citet[NEWH$\alpha$ $z=0.84$,][]{Ly2011},
\citet[HiZELS $z=0.84$,][]{Sobral2013}, and \citet[$z=2.2$,][]{Hayes2010}.

\begin{figure}[!h]
    \centering
    \includegraphics[width=5in]{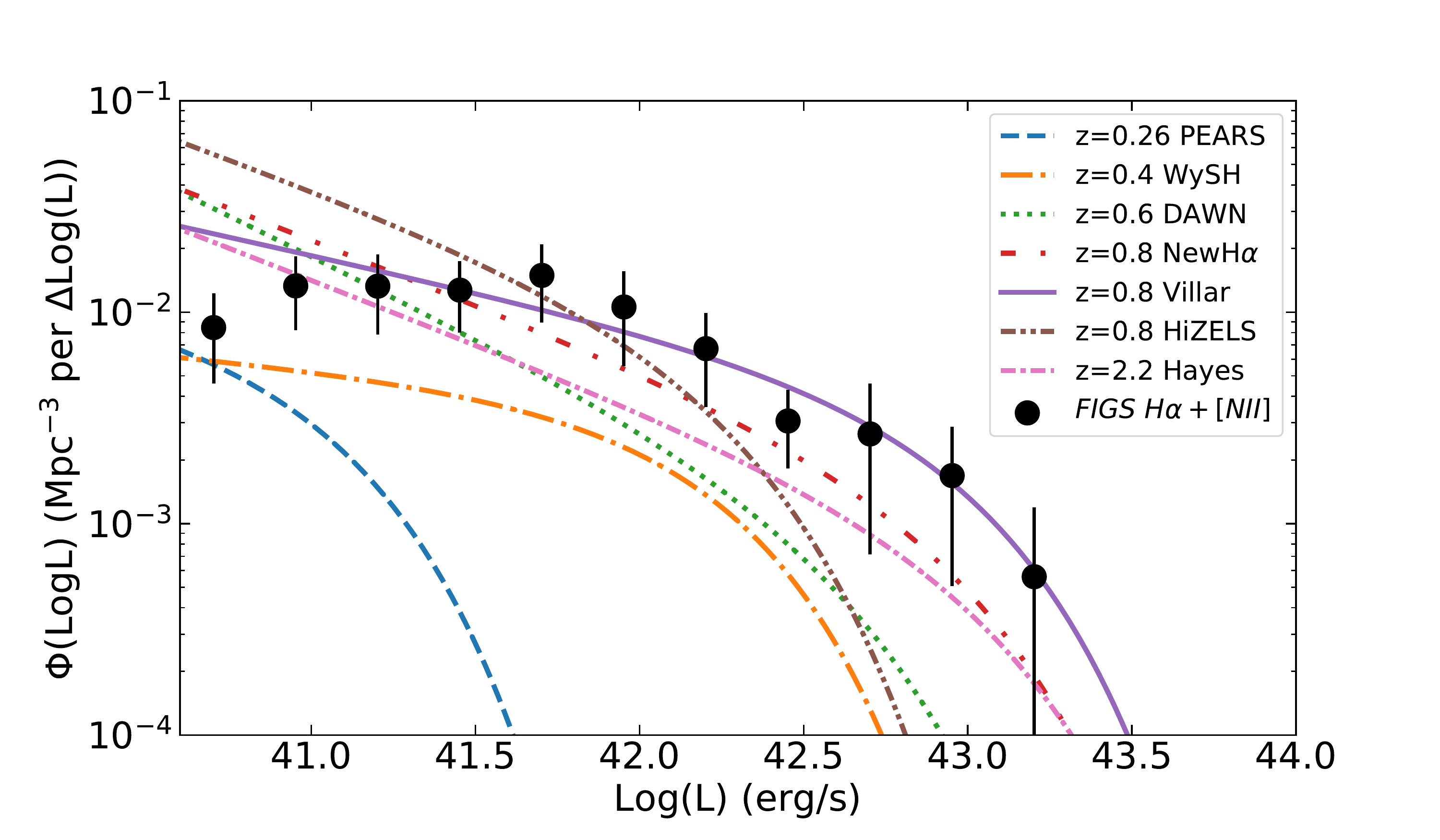}
    \caption{FIGS \Hag\ emission line luminosity function at z=0.75, completeness and extinction corrected. We also show the H$\alpha$ luminosity function from the DAWN survey at z=0.6 from \citet{Gonzalez2018} and references therein.}
    \label{fig:Halum}
\end{figure}

In order to do so, we must recompute the luminosity function for the FIGS H$\alpha$\ galaxies by taking into account the  total integrated luminosity for each individual galaxy. We can estimate the total integrated  \Hag\ flux by simply measuring it from the FIGS spectra of these EM2D-selected objects,  which we have described in \citet{Pirzkal2017}. This measured line flux is a better estimate of the integrated flux for the entire object in cases of diffuse emission, or when multiple emission-line regions are present. We applied a completeness correction factor based on the value of the brightest emission line region. We also added the luminosity-dependent dust attenuation law from \citet{Hopkins2001}. The result is shown in Figure \ref{fig:Halum}. As this Figure shows, this results is most consistent with the results of \citet{Villar2008}, despite the widely different selection method and survey areas and less consistent with the results of \citet{Sobral2013} as we detect a larger number of high-luminosity sources.

\begin{figure}[!h]
    \centering
    \includegraphics[width=6in]{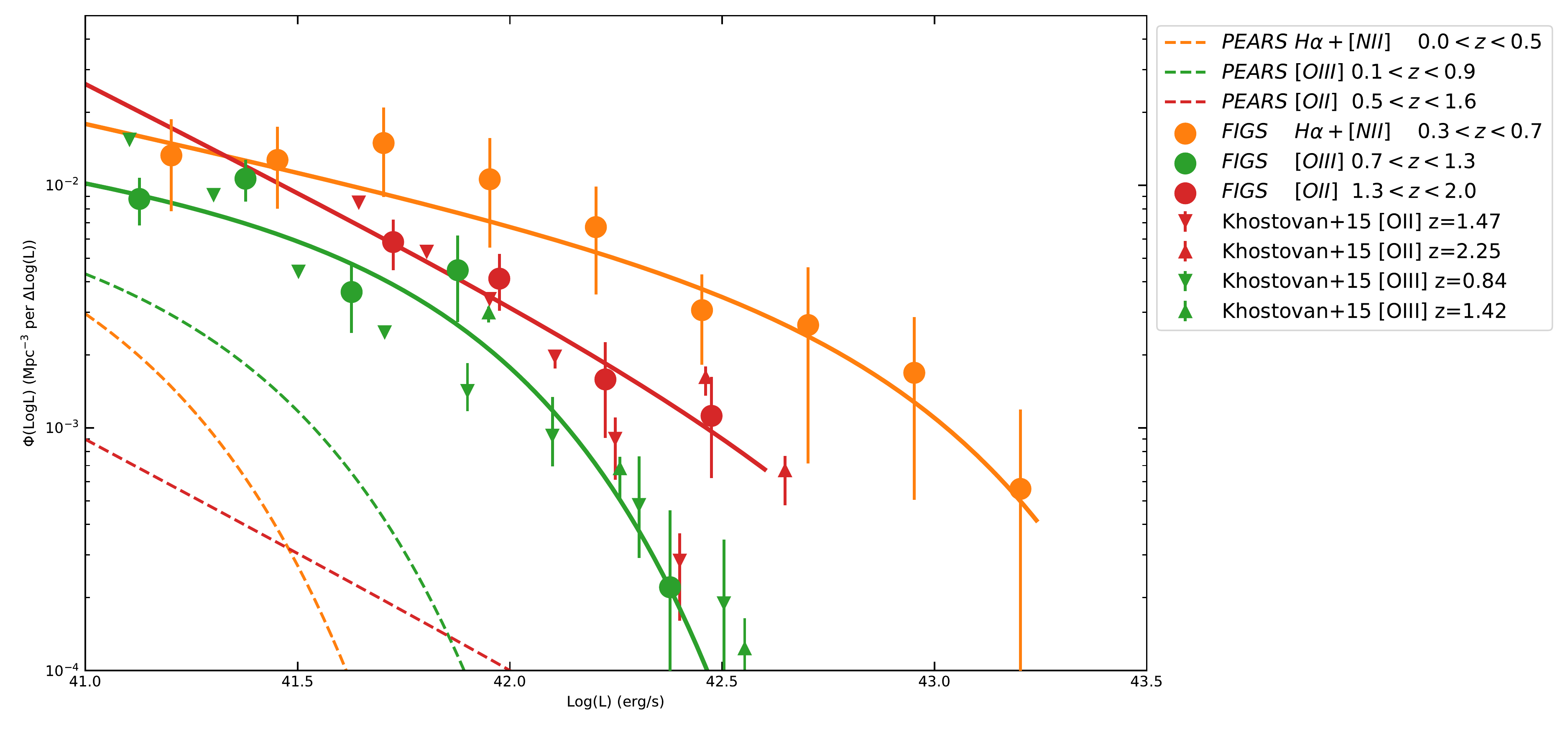}
    \caption{Luminosity functions derived for the FIGS survey for \Hag, \OIII5007, and \OIIg. We also plot the luminosity functions from \citet{Pirzkal2013}, derived at lower redshifts for the PEARS project (dash lines), and from \citet[][triangle symbols]{Khostovan2015}}
    \label{fig:lums}
\end{figure}

The luminosity functions for the \Hag,\OIII5007, and \OIIg\ emission line galaxies are shown in Figure \ref{fig:lums}. This Figure was derived by estimating the total integrated \Hag\, [OIII], and [OII] observed flux from each host galaxy and is meant to show the observed densities of these targets, accounting only for instrumental completeness and not applying dust correction. For comparison, we also plot the luminosity functions from \citet{Pirzkal2013}, derived for ELGs selected in a similar manner but at lower redshifts. While the two surveys show similar object densities over the overlapping redshift ranges of the PEARS and FIGS surveys, the FIGS survey detected significantly brighter line emission. This is consistent with stronger star formation in galaxies at an epoch that is closer to the peak of star formation history at $1.5<z<2.5$.  We also show measurements from \citet{Khostovan2015} and show good agreement with these observations too.

% \begin{figure}[!h]
%     \centering
%     \includegraphics{EM2D_lums}
%     \caption{Emission line luminosity function derived for the FIGS survey. The vertical lines show the limiting luminosity of the FIGS survey for an emission line with an observed flux of ${\rm 1\times 10^{-17} erg/s/cm^2}$ at the highest redshift reached by the FIGS survey. We also plot (blue) the luminosity function of H$\alpha$\ sources from \cite{Ly2010} for comparison and show good agreement between both measurements}
%     \label{fig:lums}
% \end{figure}

% Dropbox/FIGS_spz/Lum Function FIGS All V9 Broad c-Copy 1.ipynb

\section{Conclusion}
In order to study star formation within a wide range of galaxies, one first needs to establish a proper sample of objects, with as little selection bias as possible. Being able to detect multiple emission-line regions within galaxies, or detect extended and diffuse emission-line regions, was the goal of this paper.

We showed how we used our EM2D technique to generate a catalog of ELGs over a  wide range of redshifts by applying this method to the very deep WFC3 G102 slitless spectroscopic data obtained from the FIGS survey. There are several advantages to using the EM2D method. First, emission lines are detected independently of their host galaxies. This allows for the detection of emission lines with large EWs. Second, the exact location (to within about 0.5 WFC3 pixel, or $0.^{''}06$) of the source of the emission line can be determined, yielding both more accurate wavelength estimates and better spectroscopic redshifts. Third, strong multiple emission lines, such as \OIIg, \OIIIg + \Hb4861, and \Hag\ are detected independently, providing a check on the self-consistency of the emission-line region detections. Fourth, multiple-emission line regions, or diffuse extended emission-line regions, can be identified. 
Building on this selection process, we have shown how 2D emission line maps can be created. A detailed modeling of the physical parameters of these galaxies will be presented in the next paper of the series.
We found that a large fraction of FIGS galaxies  show signs of star formation occurring in multiple regions. We also identified a significant fraction of these objects to have extended or diffuse star formation, as opposed to bulge-dominated emission. We did not identify naked emission lines without continuum, but identified several very high EW sources, including one Ly-$\alpha$\ source at z=7.5. Overall, approximately 20\% of our [OIII] sample has a star forming region with ${\rm EW>300}$\AA. Finally, we showed that to derive accurate spectroscopic redshifts using slitless observations, one should rely on observations taken at multiple position angles and on methods such as EM2D. This should prove relevant to the planning of observations using future missions using slitless spectroscopic modes such as JWST (NIRISS and NIRCAM) and WFIRST.

\section*{Acknowledgements}
This work is based on observations made with the NASA/ESA Hubble Space Telescope, obtained [from the Data Archive] at the Space Telescope Science Institute, which is operated by the Association of Universities for Research in Astronomy, Inc., under NASA contract NAS 5-26555. These observations are associated with program \#13779 \\
Support for program \#13779 was provided by NASA through a grant from the Space Telescope Science Institute, which is operated by the Association of Universities for Research in Astronomy, Inc., under NASA contract NAS 5-26555.\\
E.C.L., J.C. and S.C. acknowledge support from the European Research Council (ERC) via an Advanced Grant under grant agreement no. 321323- NEOGAL. 
The authors would like to thank R.T. Gatto for discussions that provided invaluable assistance in developing the themes presented in this manuscript. 
AC acknowledges the grants ASI n.I/023/12/0 "Attività relative alla fase B2/C per la missione Euclid" and PRIN MIUR 2015 "Cosmology and Fundamental Physics: illuminating the Dark Universe with Euclid."

\newpage
\clearpage


\begin{thebibliography}{}
\bibitem[Aller(1942)]{Aller1942} Aller, L.~H.\ 1942, \apj, 95, 52 

\bibitem[Atek et al.(2010)]{Atek2010} Atek, H. et al. 2010, \apj, 723, 104
\bibitem[Avila et al.(2015)]{Avila2015} Avila, R.~J., Hack, W., Cara, M., et al.\ 2015, Astronomical Data Analysis Software an Systems XXIV (ADASS XXIV), 495, 281 
\bibitem[Boroson et al.(1993)]{Boroson1993} Boroson, T.~A., Salzer, J.~J., \& Trotter, A.\ 1993, \apj, 412, 524 
\bibitem[Cardamone et al.(2009)]{Cardamone2009} Cardamone, C., Schawinski, K., Sarzi, M., et al.\ 2009, \mnras, 399, 1191 
\bibitem[Chevallard \& Charlot(2016)]{Chevallard2016} Chevallard, J., \& Charlot, S.\ 2016, \mnras, 462, 1415 
\bibitem[Bertin et al.(1996)]{Bertin1996}Bertin, E., \& Arnouts, S.\ 1996, \aaps, 117, 393 
\bibitem[Cowie et al.(1996)]{Cowie1996} Cowie, L.~L., Songaila, A., Hu, E.~M., \& Cohen, J.~G.\ 1996, \aj, 112, 839 
\bibitem[Dale et al.(2010)]{Dale2010} Dale, D.~A., Barlow, R.~J., Cohen, S.~A., et al.\ 2010, \apjl, 712, L189 
\bibitem[Djorgovski et al.(1985)]{Djorgovski1985} Djorgovski, S., Spinrad, H., McCarthy, P., \& Strauss, M.~A.\ 1985, \apjl, 299, L1 
\bibitem[Ester et al. (1996)]{Ester1996} Ester, M., Kriegel, H.P., Sander, J., Xu, X. 1996, ``A density-based algorithm for discovering clusters in large spatial databases with noise,'' in Proceedings of the Second International Conference on Knowledge Discovery and Data Mining  AAAI Press. pp. 226–231
\bibitem[Giavalisco et al.(2004)]{Giavalisco2004} Giavalisco, M., Ferguson, H.~C., Koekemoer, A.~M., et al.\ 2004, 
\apjl, 600, L93 
\bibitem[Gonzalez(2018)]{Gonzalez2018} Gonzalez, A., Rhoads, J. E., Malhotra, S. et al. in prep
\bibitem[Hayes et al.(2010)]{Hayes2010} Hayes, M., Schaerer, D., \& {\"O}stlin, G.\ 2010, \aap, 509, L5 
\bibitem[Hopkins et al.(2001)]{Hopkins2001} Hopkins, A.~M., Connolly, A.~J., Haarsma, D.~B., \& Cram, L.~E.\ 2001, \aj, 122, 288 
\bibitem[Hopkins(2004)]{Hopkins2004} Hopkins, A.~M.\ 2004, \apj, 615, 209 
\bibitem[Kennicutt (1998)]{Kennicutt1998} Kennicutt, R.~C., Jr.\ 1998, \araa, 36, 189 
\bibitem[Kewley \& Dopita(2002)]{Kewley2001} Kewley, L.~J., \& Dopita, M.~A.\ 2002, \apjs, 142, 35 
\bibitem[K{\"u}mmel et al.(2009)]{Kummel2009} K{\"u}mmel, M., Walsh, J.~R., Pirzkal, N., Kuntschner, H., \& Pasquali, A.\ 2009, \pasp, 121, 59 
\bibitem[Larson et al.(2017)]{Larson2018} Larson, R.~L., Finkelstein, S.~L., Pirzkal, N., et al.\ 2018, \apj, 858, 94 
\bibitem[Luo et al.(2017)]{Luo2017} Luo, B., Brandt, W.~N., Xue, Y.~Q., et al.\ 2017, \apjs, 228, 2 
\bibitem[Ly et al.(2011)]{Ly2011} Ly, C., Lee, J.~C., Dale, D.~A., et al.\ 2011, \apj, 726, 109 
\bibitem[Osterbrock \& Ferland(2006)]{Osterbrock2006} Osterbrock, D.~E., \& Ferland, G.~J.\ 2006, Astrophysics of gaseous nebulae and active galactic nuclei, 2nd.~ed.~by D.E.~Osterbrock and G.J.~Ferland.~Sausalito, CA: University Science Books, 2006, 
\bibitem[Khostovan et al.(2015)]{Khostovan2015} Khostovan, A.~A., Sobral, D., Mobasher, B., et al.\ 2015, \mnras, 452, 3948 
\bibitem[Markarian (1967)]{Markarian1967} Markarian, B.~E.\ 1967, Astrofizika, 3, 55 
\bibitem[McCarthy et al.(1999)]{McCarthy1999} McCarthy, P.~J., Yan, L., Freudling, W., et al.\ 1999, \apj, 520, 548 
\bibitem[Madau et al.(1998)]{Madau1998} Madau, P., Pozzetti, L., \& Dickinson, M.\ 1998, \apj, 498, 106 
\bibitem[Madau \& Dickinson (2014)]{Madau2014} Madau, P., \& Dickinson, M.\ 2014, \araa, 52, 415 
\bibitem[Mayall (1936)]{Mayall1936} Mayall, N.~U.\ 1936, \pasp, 48, 14 
\bibitem[McGaugh(1991)]{McGaugh1991} McGaugh, S.~S.\ 1991, \apj, 380, 140 
\bibitem[Momcheva et al.(2016)]{Momcheva2016} Momcheva, I.~G., Brammer, G.~B., van Dokkum, P.~G., et al.\ 2016, \apjs, 225, 27 
\bibitem[Nagao et al.(2006)]{Nagao2006} Nagao, T., Maiolino, R., \& Marconi, A.\ 2006, \aap, 459, 85 
\bibitem[Oke \& Gunn(1983)]{Oke1983} Oke, J.~B., \& Gunn, J.~E.\ 1983, \apj, 266, 713 

\bibitem[Pasquali et al.(2003)]{Pasquali2003} Pasquali, A., Pirzkal, N., Walsh, J.~R., et al.\ 2003, Astronomy, Cosmology and Fundamental Physics, 471 
\bibitem[Pirzkal et al.(2001)]{Pirzkal2001} Pirzkal, N., Pasquali, A., \& Demleitner, M.\ 2001, Space Telescope European Coordinating Facility Newsletter, 29, 5 

\bibitem[Planck Collaboration et al. (2015)]{Planck2015}Planck Collaboration, Ade, P. A. R., Aghanim, N., et al. 2014, \aap, 571, A16

\bibitem[Pirzkal et al.(2004)]{Pirzkal2004} Pirzkal, N., et al. 2004, \apjs, 154, 501
\bibitem[Pirzkal et al.(2009)]{Pirzkal2009} Pirzkal, N., et al. 2009, \apj, 695, 1591
\bibitem[Pirzkal et al.(2013)]{Pirzkal2013} Pirzkal, N., Rothberg, B., Ly, C., et al.\ 2013, \apj, 772, 48 
\bibitem[Pirzkal et al.(2016)]{Pirzkal2016} Pirzkal, N., Ryan, R., \& Brammer, G.\ 2016, Space Telescope WFC Instrument Science Report, 15 
\bibitem[Pirzkal et al.(2017)]{Pirzkal2017} Pirzkal, N., Malhotra, S., Ryan, R.~E., et al.\ 2017, \apj, 846, 84 
\bibitem[Rubin et al.(1970)]{Rubin1970} Rubin, V.~C., Ford, W.~K., Jr., \& D'Odorico, S.\ 1970, \apj, 160, 801 
\bibitem[Rubin et al.(1972)]{Rubin1972} Rubin, V.~C., Krishna Kumar, C., \& Ford, W.~K., Jr.\ 1972, \apj, 177, 31 
\bibitem[Smith (1975)]{Smith1975} Smith, M.~G.\ 1975, \apj, 202, 591 
\bibitem[Sobral et al.(2013)]{Sobral2013} Sobral, D., Smail, I., Best, P.~N., et al.\ 2013, \mnras, 428, 1128
\bibitem[Storey \& Zeippen(2000)]{Storey2000} Storey, P.~J., \& Zeippen, C.~J.\ 2000, \mnras, 312, 813 
\bibitem[Straughn et al.(2008)]{Straughn2008} Straughn, A.~N., Meurer, G.~R., Pirzkal, N., et al.\ 2008, \aj, 135, 1624 
\bibitem[Straughn et al.(2009)]{Straughn2009} Straughn, A.~N., Pirzkal, N., Meurer, G.~R., et al.\ 2009, \aj, 138, 1022 
\bibitem[Tilvi et al.(2016)]{Tilvi2016} Tilvi, V., Pirzkal, N., Malhotra, S., et al.\ 2016, \apjl, 827, L14 
\bibitem[Treu et al.(2015)]{Treu2015} Treu, T., Schmidt, K.~B., Brammer, G.~B., et al.\ 2015, \apj, 812, 114 \bibitem[Xia et al.(2011)]{Xia2011} Xia, L., Malhotra, S., Rhoads, J., et al.\ 2011, \aj, 141, 64 
\bibitem[Xu et al.(2007)]{Xu2007} Xu, C., Pirzkal, N., Malhotra, S., et al.\ 2007, \aj, 134, 169 
\bibitem[Villar et al.(2008)]{Villar2008} Villar, V., Gallego, J., P{\'e}rez-Gonz{\'a}lez, P.~G., et al.\ 2008, \apj, 677, 169
\bibitem[Villforth et al.(2010)]{Villforth2010} Villforth, C., Koekemoer, A.~M., \& Grogin, N.~A.\ 2010, \apj, 723, 737 
\bibitem[Xue et al.(2016)]{Xue2016} Xue, Y.~Q., Luo, B., Brandt, W.~N., et al.\ 2016, \apjs, 224, 15 
\bibitem[Yang et al.(2017)]{Yang2017} Yang, H., Malhotra, S., Rhoads, J.~E., \& Wang, J.\ 2017, \apj, 847, 38 
\end{thebibliography}
\end{document}